\documentclass[a4paper,usenatbib]{mnras}
\usepackage{graphicx}
\usepackage[fleqn]{amsmath}
\usepackage[T1]{fontenc}
\usepackage{aecompl}
\usepackage{times}
\usepackage{ae,aecompl}
\usepackage{amssymb,amsfonts,hyperref,color}

\newcommand\lsun{\rm L_{\odot}}
\newcommand\msunyr{\rm M_{\odot}\,yr^{-1}}
\newcommand\be{\begin{equation}}
\newcommand\en{\end{equation}}

\newcommand\etal{{\rm et al}.\ }

\title[Radiation MHD Simulations for FU Ori]{Global 3-D Radiation Magnetohydrodynamic Simulations for FU Ori's Accretion Disk and Observational Signatures of  Magnetic Fields}

\author[Z.~Zhu, Y.-F.~Jiang, J.~Stone]{
Zhaohuan Zhu$^{1}$\thanks{E-mail: zhaohuan.zhu@unlv.edu},  Yan-Fei Jiang$^{2}$, James M. Stone$^{3}$ \\
$^{1}$Department of Physics and Astronomy, University of Nevada, Las Vegas, 4505 S.~Maryland Parkway, Las Vegas, NV~89154, USA\\
$^{2}$Center for Computational Astrophysics, Flatiron Institute, New York, NY 10010, USA\\
$^{3}$Institute for Advance Study, 1 Einstein Drive, Princeton, NJ, 08540, USA\\}
\date{In original form \today}

\begin{document}
\label{firstpage}
\pagerange{\pageref{firstpage}--\pageref{lastpage}} \pubyear{2019}
\maketitle

\begin{abstract}
FU Ori is the prototype of FU Orionis systems which are outbursting protoplanetary disks. Magnetic fields in FU Ori's accretion disks have previously been detected using spectropolarimetry observations for Zeeman effects. 
We carry out global radiation ideal MHD simulations to study FU Ori's inner accretion disk. We find that (1) when the disk is threaded by vertical magnetic fields, most accretion occurs in the magnetically dominated atmosphere at z$\sim$R, similar to the ``surface accretion'' mechanism in previous locally-isothermal MHD simulations. (2) A moderate disk wind is launched  in the vertical field simulations with a terminal speed of $\sim$300-500 km/s and a mass loss rate of  1-10\% the disk accretion rate, which is consistent with observations. Disk wind fails to be launched in simulations with net toroidal magnetic fields. (3) The disk photosphere at the unit optical depth can be either in the wind launching region or the accreting surface region. Magnetic fields have drastically different directions and magnitudes between these two regions. Our fiducial model agrees with previous optical Zeeman observations regarding both the field directions and magnitudes. On the other hand, simulations indicate that future Zeeman observations at near-IR wavelengths or towards other FU Orionis systems may reveal very different magnetic field structures. (4) Due to energy loss by the disk wind, the disk photosphere temperature is lower than that predicted by the thin disk theory, and the previously inferred disk accretion rate may be lower than the real accretion rate by a factor of  $\sim$2-3.   
\end{abstract}


\section{Introduction}
Accretion disks have been observed in a wide range of astrophysical systems, ranging from around low mass stars \citep{Hartmann2016} to around compact objects and
supermassive black holes \citep{Begelman1984}. The accretion process not only helps to build the central object, but the released radiation energy allows us to identify and study the central
object (e.g. X-ray binaries). The high resolution M87 image by the Event Horizon Telescope \citep{Event2019} is an excellent example that we can constrain the properties of black holes by studying their surrounding accretion disks. 

The leading theory to explain the accretion process involves magnetic fields, especially for sufficiently ionized disks\footnote{In poorly ionized disks where the non-ideal MHD effects become 
important, hydrodynamical processes may also play an important role in disk accretion \citep{Turner2014a}. }.  Magnetic fields can drive turbulence through the
magnetorotational instability (MRI; \citealt{BalbusHawley1991, BalbusHawley1998}) or/and launch disk winds through the
magnetocentrifugal effect in non-relativistic disks \citep{BlandfordPayne1982}. The strengths of both MRI turbulence and disk winds depend on the field strength. Normally turbulence and wind are more prominent in systems having stronger magnetic fields \citep{Hawley1995}.

Despite the importance of magnetic fields, the observational evidences for magnetic fields  in accretion disks remain to be scarce. 
The collimated jets/outflows provide some indirect evidences of magnetic fields since the confinement of jets may require the presence of
magnetic fields \citep{Pudritz2007, Frank2014}. Another indirect evidence is from magnetic field measurements from meteorites. Paleomagnetic measurements
by \cite{Fu2014} suggest that Semarkona meteorites were magnetized to 0.54 G in the solar nebulae. 

The most direct evidence of magnetic fields in accretion disks comes
from Zeeman splitting of atomic or molecular lines. Current Zeeman measurements of molecular lines using ALMA \citep{vlemmings2019} have only placed upper limits on the field strength ($<$  30 mG).  So far, the only direct measurement of magnetic fields in accretion disks is the detection of Zeeman splitting of atomic lines coming from the inner disk of FU Ori \citep{Donati2005}.  

FU Ori is the prototype of
FU Orionis systems: a small but remarkable class of variable young stellar objects that undergo outbursts in optical light of 5 magnitudes or more \citep{Herbig1977}. 
While the outburst has a fast rise time ($\lesssim$ 1-10 yr), the decay timescale ranges from decades to centuries \citep{audard2014,Connelley2018}. 
Although more FU Orionis outbursts have been discovered
recently thanks to large-scale all-sky surveys (e.g. \citealt{Semkov2010, Kraus2016, Kospal2017, Hillenbrand2018}), the occurrence rate of these objects among young stars is still illusive \citep{Hillenbrand2015, Scholz2013} with rates ranging from less than 1 outburst per young star to more than tens of outbursts per young star. 

Such intense outbursts are due to the sudden increase of the protostellar disk's accretion rate from $\sim10^{-8}\msunyr$ (Class I-II rates) 
to $\sim10^{-4}\msunyr$ \citep{HartmannKenyon1996}. The strong accretion is accompanied by the strong disk wind \citep{Calvet1993,Milliner2019}.
Although the outburst triggering mechanism is not clear\footnote{Current theory includes fragmented clumps \citep{VorobyovBasu2006}, spiral arms from gravitational instability \citep{Armitage2001,Zhu2009b,Martin2012,Bae2014,Kadam2019}, or binary interaction \citep{Bonnell1992}.}, the inner disks ($\lesssim$1 au) during the outbursts are hot enough ($\sim$6000 K, \citealt{Zhu2007}) to be sufficiently ionized and 
MRI should operate in these disks.
Since these inner disks with $\sim 100\lsun$ are much brighter than the central stars and all the light we see are from these accretion disks, FU Orionis systems are ideal places to study accretion physics. 

Taking advantage of many atomic lines available in these systems, \cite{Donati2005} have used the high-resolution spectropolarimeter to
detect signals of Zeeman splitting in FU Ori. By splitting the circular polarization signal into symmetric and antisymmetric components, they
constrain the magnetic fields in both the azimuthal and radial directions. 
Assuming that the disk's rotational axis is 60$^o$ inclined with respect to our line of sight,  
their best fit model suggests that 
the vertical component of the fields is $\sim$ 1 kG at 0.05 au and points towards the observer, while the azimuthal component (about half as strong) points in a direction opposite to the orbital rotation. 

In spite of these stringent observational constraints, theoretical work still lacks behind and its connection with observations has not been established. To study FU Ori using theoretical numerical simulations,
high enough numerical resolution is necessary for capturing MRI, while a large simulation domain is needed to study the disk wind. 
Only recently, with the newly developed Athena++ code which has both mesh-refinement and the special polar boundary condition, we can simulate the whole 4$\pi$ sphere around the central object
with enough resolution to capture MRI \citep{ZhuStone2018}. 
Besides magnetic fields, radiative transfer is also crucial for understanding FU Ori's inner accretion disk. For example, thermal instability was previously suggested to explain FU Ori's outburst \citep{BellLin1994}. Although local shearing box MHD simulations with radiative transfer \citep{Hirose2014} do not support the thermal instability theory for FU Ori outbursts \citep{Hirose2015},
the disk's thermal structure is still important for both the accretion physics \citep{Zhu2009c} and the boundary layer physics \citep{KleyLin1999} .
Furthermore, radiative transfer is important for making connections with observations (e.g. understanding the physical condition at the disk's photosphere).

Thus, in this work, we include radiative transfer in the global MHD disk simulations to study the accretion structure of FU Ori's inner disk. We will also compare our simulations with
 previous Zeeman magnetic field observations and disk wind observations. 
In Section 2, the theoretical framework for energy transport in accretion disks is presented. 
We will describe our numerical method in Section 3. The results are presented in Section 4. After connecting  with observations and a short discussion
 in Section 5, the paper is concluded in Section 6.

\section{Theoretical Framework}
Angular momentum transport and energy transport are two important aspects of the accretion disk theory. 
Angular momentum transport is essential for the mass buildup of the central object, while energy transport is crucial for revealing disk properties using observations.  
Previously in \cite{ZhuStone2018}, we have done detailed analyses on angular momentum transport for disks
threaded by net vertical magnetic fields. In this work, we will focus on energy transport in accretion disks. 

The fluid equations with both magnetic and radiation fields are
\begin{eqnarray}
\frac{\partial\rho}{\partial t}+\nabla\cdot\left(\rho{\bf v}\right)&=&0\nonumber\\
\frac{\partial\rho{\bf v}}{\partial t}+\nabla\cdot\left(\rho{\bf v}{\bf v}-{\bf B}{\bf B}+{\sf P^*}+{\sf \sigma}\right)&=&-{\bf S_{r}}({\bf P})+{\bf F}\nonumber\\
\frac{\partial E}{\partial t}+\nabla\cdot\left[\left(E+P^{*}\right){\bf v}-{\bf B}\left({\bf B}\cdot{\bf v}\right)+{\sf \sigma}\cdot{{\bf v}}\right]&=&-cS_{r}(E)+{\bf F}\cdot{\bf v}\nonumber\\
\frac{\partial {\bf B}}{\partial t}-\nabla\times\left({\bf v}\times{\bf B}\right)&=&0\,,\label{eq:fullequation}
\end{eqnarray}
where $E=E_{g}+\rho v^2/2+B^2/2$ is the total gas energy density, $E_{g}=P/(\gamma-1)$ is the internal energy, ${\sf P^*}\equiv (P+B^2/2){\sf I}$ is the pressure tensor (with ${\sf I}$ the unit tensor), and ${\bf F}$ is the external force (e.g. gravity).
We also include the dissipation tensor ${\sf \sigma}$ in the equations. Although dissipation is not explicitly added in the simulations, shock dissipation is implicitly included in the Riemann solver, and
dissipation terms are important for the energy analysis. 
The radiation equations are
\begin{eqnarray}
\frac{\partial E_{r}}{\partial t}+\nabla\cdot {\bf F_{r}}=cS_{r}(E)\\
\frac{1}{c^2}\frac{\partial {\bf F_r}}{\partial t}+\nabla\cdot{\sf P_{r}}=\bf{S_{r}({\bf P})}\,,
\end{eqnarray}
where the radiation flux ${\bf F_{r}}$ and the radiation energy density $E_{r}$ are Eulerian variables, and they are related to the co-moving flux
${\bf F_{r,0}}$ through ${\bf F_{r,0}}={\bf F_{r}}-({\bf v}E_{r}+{\bf v}\cdot {\sf P_r})$. The radiation pressure tensor
${\sf P_r}$ is related to the energy density though the variable Eddington tensor ${\sf P_r}={\sf f}E_{r}$. The source terms $cS_{r}(E)$ and $\bf{ S_{r}}({\bf P})$
are given in \cite{Jiang2013}. 

To study the energy budget, it is also helpful to write the equation for the gas' internal energy density. 
The kinetic and magnetic energy equation is
\begin{eqnarray}
&&\frac{\partial }{\partial t}\left(\frac{\rho v^2}{2}+\frac{B^2}{2}\right)\nonumber\\
&+&\nabla\cdot\left[{\bf v} \left(\frac{\rho v^2}{2}\right)-{\bf B}\left({\bf B}\cdot{\bf v}\right)+({\sf P^*}+{\sf \sigma})\cdot {\bf v}\right]\nonumber\\
&-&\left(P-\frac{B^2}{2}\right)\nabla\cdot{\bf v}+ 
\left({\bf v}\cdot \nabla\right)\frac{B^2}{2}-\left({\sf \sigma\cdot\nabla}\right)\cdot{\bf v}\nonumber\\
&=&-{\bf v}\cdot \bf{S_{r}({\bf P})}+{\bf F}\cdot{\bf v}\,,
\end{eqnarray}
so that the internal energy density is
\begin{equation}
\frac{\partial E_g}{\partial t}+\nabla\cdot\left(E_g{\bf v}\right)+P\nabla\cdot{{\bf v}}+\left({\sf \sigma}\cdot\nabla\right)\cdot{\bf v}=-cS_r(E)+{\bf v}\cdot{\bf S_{r}(P)}\,,\label{eq:internal}
\end{equation}
which suggests that the change of the internal energy is due to the $Pdv$ work, the dissipation, and radiative transport. 

We can use either the equation for the total energy (Equation \ref{eq:fullequation}) or the equation for the internal energy (Equation \ref{eq:internal}) to derive the disk luminosity.
Here,  we rewrite the total energy equation as
\begin{equation}
\frac{\partial E}{\partial t}+\nabla\cdot {\bf A}=-Q_{cool}+ \bf{F}\cdot\bf{v}\,,\label{eq:energy}
\end{equation}
where ${\bf A}=(E+P^{*})\bf{v}-{\bf B}({\bf B}\cdot{\bf v})$, and $Q_{cool}$ is the radiative cooling rate. ${\bf A}$ can also be rewritten as
\begin{equation}
{\bf A}=(\frac{\gamma}{\gamma-1}P+\frac{1}{2}\rho v^2){\bf v}+{\bf B}\times({\bf v}\times{\bf B})
\end{equation}
using vector identities.

We will first review the thin disk theory under the cylindrical coordinate system and then we will write similar equations under the spherical-polar coordinate system that has been adopted in our simulations.
The perturbed equation for the angular momentum under the cylindrical coordinate system can be written as
\begin{eqnarray}
\frac{\partial \langle\rho\delta v_{\phi}\rangle}{\partial t}&=&-\frac{1}{R^2}\frac{\partial (R^2  \langle T_{R\phi}\rangle)}{\partial R}-\frac{\langle\rho v_{R}\rangle}{R}\frac{\partial Rv_{K}}{\partial R}\nonumber\\
&&-\frac{\partial \langle T_{\phi z}\rangle}{\partial z}-
\langle\rho v_{z}\rangle\frac{\partial  v_{K}}{\partial z}\,,\label{eq:angcyl}
\end{eqnarray}
where 
\begin{align}
T_{R\phi}\equiv \rho v_{R}\delta v_{\phi}-B_{R}B_{\phi}\nonumber\\
T_{\phi z}\equiv \rho v_{z}\delta v_{\phi}-B_{z}B_{\phi}\,,
\end{align}
and $\langle\rangle$ denotes that the quantity has been averaged in the azimuthal ($\phi$) direction.
Assuming a steady state, we have
\begin{equation}
\frac{\dot{M}}{2\pi}\frac{\partial Rv_{K}}{\partial R}=\frac{\partial (R^2  \langle T_{R\phi}\rangle)}{\partial R}+R^2\frac{\partial \langle T_{\phi z}\rangle}{\partial z}+R^2\langle\rho v_{z}\rangle\frac{\partial  v_{K}}{\partial z}\label{eq:stresscyl}\,,
\end{equation}
where $\dot{M}\equiv -2\pi R\langle\rho v_{R}\rangle$. Thus, the accretion is driven by the $T_{R\phi}$ stress within the disk or the $T_{\phi z}$ stress at the disk surface.
If we assume that $\dot{M}$ is a constant along $R$,
we have
\begin{equation}
\langle T_{R\phi}\rangle=\frac{\dot{M}v_{K}}{2\pi R}-\frac{C}{R^2}-\frac{1}{R^2}\int  R^2\left(\frac{\partial \langle T_{\phi z}\rangle}{\partial z}+ \langle\rho v_{z}\rangle\frac{\partial  v_{K}}{\partial z}\right)dR\,.
\end{equation}

The energy Equation (Equation \ref{eq:energy}) under the cylindrical coordinate system is
\begin{align}
\frac{\partial \langle E\rangle}{\partial t}&=-\frac{1}{R}\frac{\partial ( R  \langle A_{R}\rangle)}{\partial R}-\frac{\partial \langle A_{z}\rangle}{\partial z}-\langle Q_{cool}\rangle+\langle \bf{F}\cdot\bf{V}\rangle\,,\label{eq:encyl}
\end{align}
where the leading terms in $A_{R}$ (after removing the second-order terms) are
\begin{equation}
A_{R}=\frac{\gamma}{\gamma-1}Pv_{R}+\frac{1}{2}\rho v_{R} v_{K}^2+ v_{K}T_{R\phi}\,,
\end{equation}
and the leading terms in $A_{z}$ are
\begin{equation}
A_{z}=\frac{\gamma}{\gamma-1}Pv_{z}+\frac{1}{2}\rho v_{z} v_{K}^2+ v_{K}T_{\phi z}\,.
\end{equation}
If we ignore the pressure term in $A_{R}$, assume $v_{z}\sim0$ in $A_{z}$, and assume a steady state, we have
\begin{equation}
\langle Q_{cool}\rangle=-\frac{1}{R}\frac{\partial (   \langle -\frac{1}{4\pi}\dot{M} v_{K}^2+ Rv_{K}T_{R\phi} \rangle)}{\partial R} -\frac{\partial \langle  v_{K}T_{\phi z} \rangle}{\partial z} +\langle \bf{F}\cdot\bf{V}\rangle\,.
\end{equation}
If we plug in $T_{R\phi}$, ignore the $T_{\phi z}$ term, replace ${\bf F}$ with the gravitational force, and only consider the disk midplane, we have
\begin{equation}
2\pi\langle Q_{cool}\rangle=-\frac{1}{2}\frac{\dot{M}v_{K}^2}{R^2}+\frac{\dot{M}v_{K}^2}{R^2}-\frac{3}{2}\frac{Cv_{K}}{R^3}+ \frac{\dot{M}v_{K}^2}{R^2}\,,
\end{equation}
where the first term on the right is due to the radial derivative of the Keplerian kinetic energy flux, the second and third terms on the right are due to the radial
derivative of the $R-\phi$ stress, and the last term on the right is the release of the gravitational potential energy. With the traditional 
zero stress inner boundary condition ($C=\dot{M}R_{in}v_{K,in}$), the cooling rate is
\begin{equation}
\langle Q_{cool}\rangle=\frac{3\dot{M}v_{K}^2}{4\pi R^2}\left(1-\left(\frac{R_{in}}{R}\right)^{1/2}\right)\,.\label{eq:Qcool}
\end{equation}
After the vertical integration, this cooling rate becomes what we normally use in the thin disk approximation,
\begin{equation}
\sigma T_{eff}^4=\frac{3G\dot{M}M}{8\pi R^3}\left(1-\left(\frac{R_{in}}{R}\right)^{1/2}\right)\,.\label{eq:Qcoolteff}
\end{equation} 
If we integrate over the whole disk starting from $R_{in}$, the total cooling rate is half the release rate of the gravitational potential energy ($GM\dot{M}/2R_{in}$). 
On the other hand, far away from the central star ($R\gg R_{in}$), the cooling rate ($3\dot{M}v_{K}^2/4\pi R^2$) is actually higher than the energy release 
rate from the gravitational contraction ($\dot{M}v_{K}^2/2\pi R^2$).
The additional $\dot{M}v_{K}^2/4\pi R^2$ energy release is due to the energy transport in the radial direction. We note that the same equation can also be derived using
the internal energy equation but with an additional step to derive the dissipation term.

On the other hand, our simulated disks are very thick, and the disk photosphere flares roughly following the radial direction in the spherical grids.
Thus, we want to derive similar equations for the spherical-polar coordinate system so that we can study energy transport in our simulations. 
The perturbed angular momentum equation under the spherical-polar coordinate system is 
\begin{eqnarray}
\frac{\partial \langle\rho\delta v_{\phi}\rangle}{\partial t}=-\frac{1}{r^3}\frac{\partial ( r^3  \langle T_{r\phi}\rangle)}{\partial r}-\frac{\langle\rho v_{r}\rangle}{r}\frac{\partial rv_{K}}{\partial r}\nonumber\\
-\frac{1}{r {\rm sin}^2\theta}\frac{\partial ({\rm sin}^2\theta\langle T_{\theta\phi}\rangle)}{\partial \theta}-
\frac{\langle\rho v_{\theta}\rangle}{r {\rm sin}\theta}\frac{\partial ({\rm sin}\theta v_{K})}{\partial \theta}\,,\label{eq:angsph}
\end{eqnarray}
where 
\begin{align}
T_{r\phi}\equiv \rho v_{r}\delta v_{\phi}-B_{r}B_{\phi}\nonumber\\
T_{\theta\phi}\equiv \rho v_{\theta}\delta v_{\phi}-B_{\theta}B_{\phi}\,.
\end{align}
Assuming a steady state, we have
\begin{eqnarray}
\dot{\widetilde{M}}\frac{\partial rv_{K}}{\partial r}&=&\frac{\partial ( r^3  \langle T_{r\phi}\rangle)}{\partial r}+\frac{r^2}{{\rm sin}^2\theta}\frac{\partial ({\rm sin}^2\theta\langle T_{\theta\phi}\rangle)}{\partial \theta}\nonumber\\
&&+\frac{r^2\langle\rho v_{\theta}\rangle}{ {\rm sin}\theta}\frac{\partial ({\rm sin}\theta v_{K})}{\partial \theta}
\end{eqnarray}
where $\dot{\widetilde{M}}=-r^2\langle\rho v_{r}\rangle$. Note that this $\dot{\widetilde{M}}$ definition is different from the $\dot{M}$ definition in the cylindrical coordinate system. If we assume that $\dot{\widetilde{M}}$ is a constant along $r$, we can integrate the equation to derive
\begin{eqnarray}
\langle T_{r\phi} \rangle = \frac{\dot{\widetilde{M}}v_{K}}{r^2}-\frac{C}{r^3}-\frac{1}{r^3}\int \frac{r^2}{{\rm sin}^2\theta}\\
\left(\frac{\partial ({\rm sin}^2\theta\langle T_{\theta\phi}\rangle)}{\partial \theta}+{\rm sin}\theta\langle\rho v_{\theta}\rangle\frac{\partial ({\rm sin}\theta v_{K})}{\partial \theta}\right) dr \label{eq:Trphi}
\end{eqnarray}

The energy equation (Equation \ref{eq:energy})  under the spherical-polar coordinate system is
\small
\begin{align}
\frac{\partial \langle E\rangle}{\partial t}&=-\frac{1}{r^2}\frac{\partial ( r^2  \langle A_{r}\rangle)}{\partial r}-\frac{1}{r {\rm sin}\theta}\frac{\partial ({\rm sin}\theta\langle A_{\theta}\rangle)}{\partial \theta}-\langle Q_{cool}\rangle+\langle \bf{F}\cdot\bf{V}\rangle\,.\label{eq:ensph}
\end{align}
\normalsize

The leading terms in $A_{r}$ are 
 \begin{equation}
A_{r}=(\frac{\gamma}{\gamma-1}P+\frac{1}{2}\rho v_{K}^2+\rho v_{K}\delta v_{\phi})v_{r}-v_{K}B_{r}B_{\phi}
\end{equation}
or
\begin{equation}
A_{r}=\frac{\gamma}{\gamma-1}Pv_{r}+\frac{1}{2}\rho v_{r} v_{K}^2+ v_{K}T_{r\phi}
\end{equation}

The leading terms in $A_{\theta}$ are 
\begin{equation}
A_{\theta}=\frac{\gamma}{\gamma-1}Pv_{\theta}+\frac{1}{2}\rho v_{\theta} v_{K}^2+ v_{K}T_{\theta\phi}
\end{equation}

In \S 4.2, we will measure the energy transport due to the $A_r$ and $A_{\theta}$ terms from our simulations. 
On the other hand, in this section, we will continue the derivation by making several assumptions.
If we ignore the pressure term in $A_{r}$, assume $v_{\theta}\sim0$ in $A_{\theta}$, and assume a steady state, we have
\begin{eqnarray}
\langle Q_{cool}\rangle=-\frac{1}{r^2}\frac{\partial ( \langle -\frac{1}{2}\dot{\widetilde{M}} v_{K}^2+ r^2  v_{K}T_{r\phi} \rangle)}{\partial r}\nonumber\\
-\frac{1}{r {\rm sin}\theta}\frac{\partial ({\rm sin}\theta\langle v_{K}T_{\theta\phi} \rangle)}{\partial \theta}+\langle \bf{F}\cdot\bf{V}\rangle\,.
\end{eqnarray}
If we plug in $T_{r\phi}$ from Equation \ref{eq:Trphi} and ignore $T_{\theta\phi}$ terms, we have
\begin{equation}
\langle Q_{cool}\rangle=-\frac{1}{2}\frac{\dot{\widetilde{M}}v_{K}^2}{r^3}+\frac{\dot{\widetilde{M}}v_{K}^2}{r^3}-\frac{3 }{2}\frac{C v_{K}}{ r^4}+\frac{\dot{\widetilde{M}}v_{K}^2}{r^3}\,.\label{eq:sphcool}
\end{equation}
Thus, if we can ignore the $\theta$ direction energy advection/stress and the boundary C term, the cooling rate equals the  release rate of the
gravitational potential energy (the last term on the right) plus the radially advected energy (the first two terms on the right). Unfortunately, as will be shown in Section 4.2,
the energy advection in the $\theta$ direction and the $T_{\theta\phi}$ stress can not be ignored. Accordingly, the cooling rate is modified significantly.

\section{Method}

\begin{figure}
\includegraphics[width=3.3in]{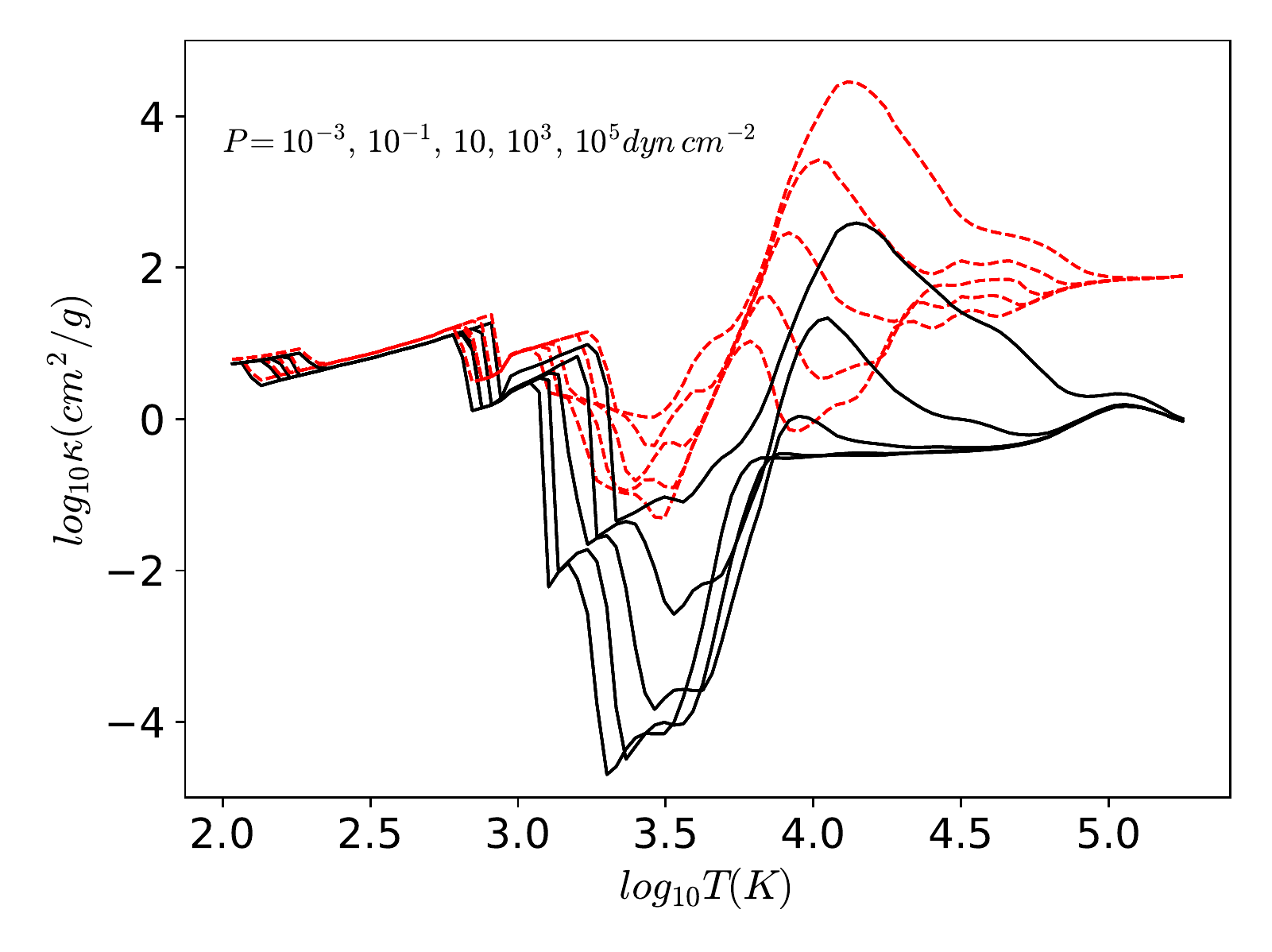}
\caption{The Rosseland mean (solid black curves) and Planck mean (red dashed curves) opacities adopted in the simulations. Different
curves represent opacities under different pressures ($10^{-3}$ to $10^5$ $dyn\ cm^{-2}$ ). Curves with overall lower values correspond to lower pressures.    }
\label{fig:rossplanck}
\end{figure}

\begin{figure*}
\includegraphics[trim=6mm 6mm 6mm 70mm, clip, width=6.6in]{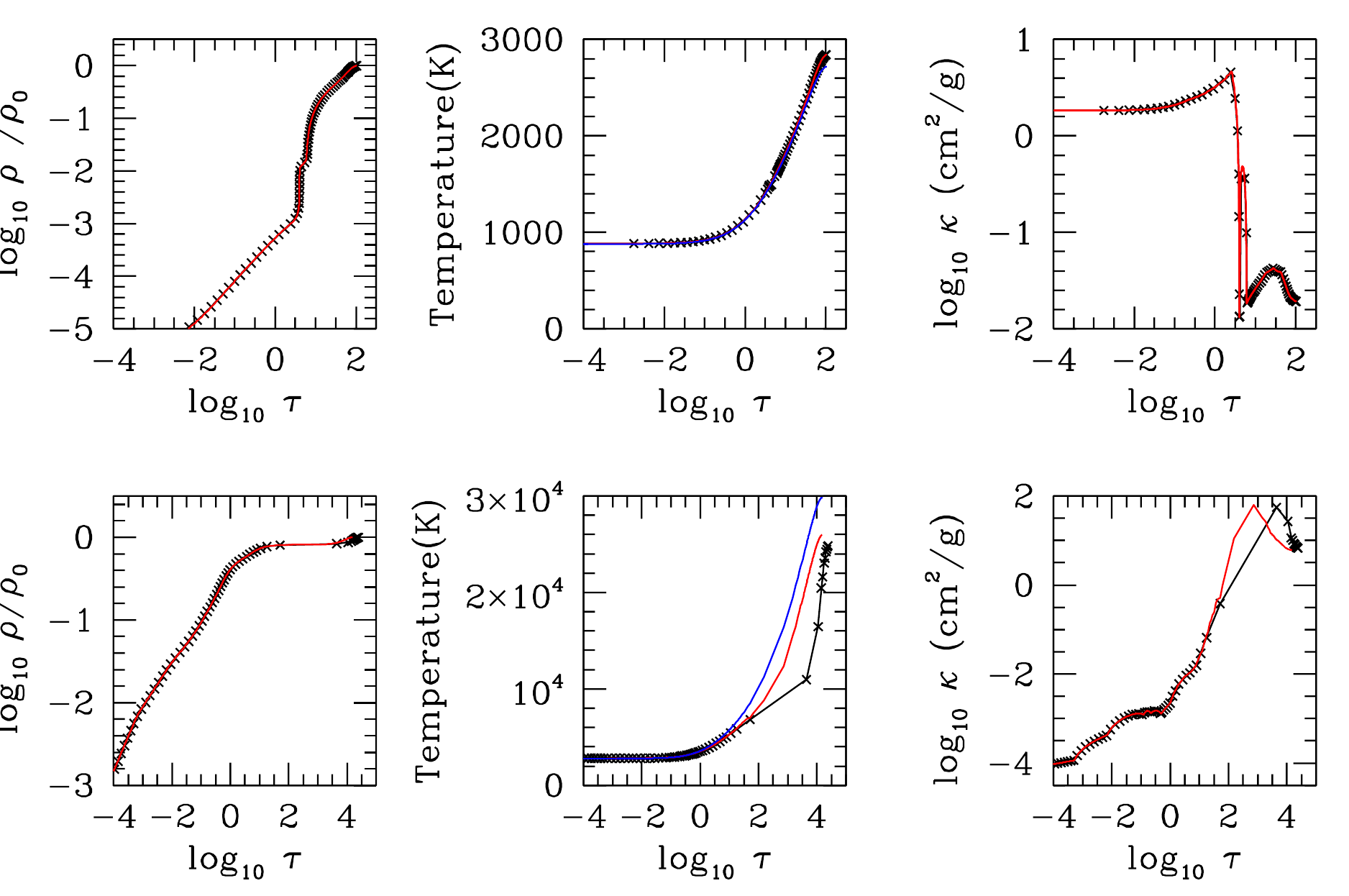}
\caption{Plane-parallel atmosphere tests for atmospheres having two different heating rates (the simulation with the lower heating rate is shown in the upper panels). 
Density, temperature, and Rosseland mean opacity at t=1000$T_0$ are shown
from the left to right panels. The black crosses and curves are results from low resolution simulations while the red curves are from the simulations with 10 times higher resolution.
The blue curves in the middle panels show the analytical temperature profiles. 
 \label{fig:test}}
\end{figure*}

We solve the magnetohydrodynamic (MHD) equations  in the ideal MHD limit using Athena++ (Stone \etal 2020, in press).
Athena++ is a newly developed grid based code using a higher-order Godunov scheme for MHD and 
the constrained transport (CT) to conserve the divergence-free property for magnetic fields. 
Compared with its predecessor Athena \citep{GardinerStone2005,GardinerStone2008,Stone2008},  Athena++ is highly optimized for 
speed and uses a flexible grid structure that enables mesh refinement, allowing global
numerical simulations spanning a large radial range. 
Furthermore, the geometric source terms in curvilinear coordinates 
(e.g. in cylindrical and spherical-polar coordinates) are specifically implemented to converse the angular momentum to machine precision. In this work, we adopt the second-order piecewise-linear method for the spatial reconstruction, the second-order Van-Leer method for the time integration, and the HLLC Riemann solver to calculate the flux. 

The time-dependent radiative transfer equation has been solved explicitly and coupled with the MHD fluid equations using the radiation module of \cite{Jiang2014}.
The general radiative transfer equation for the static fluid is
\begin{equation}
\frac{1}{c}\frac{\partial I_{\nu}}{\partial t}+ {\bf n}\cdot\nabla I_{\nu}=-(\sigma_{\nu,a}+\sigma_{\nu,s})I_{\nu}+j_{\nu}+\sigma_{\nu,s}^{eff}J_{\nu}\label{eq:iequationnu}
\end{equation}
where $I_{\nu}({\bf x},t,{\bf n})$ is the intensity at the position ${\bf x}$, time $t$ and along the direction of $\bf{n}$. $J_{\nu}=(4\pi)^{-1}\int I_{\nu}d\Omega$ and $j_{\nu}/\sigma_{\nu,a}=B_{\nu}$, while $\sigma_{\nu,a}$ and $\sigma_{\nu,s}$ are the absorption and scattering opacity at the frequency of $\nu$. However, for a fluid that is moving at $v$, additional correction terms on the order of $(v/c)$ and $(v/c)^2$ need to be added \citep{Jiang2014}.   \cite{Jiang2019a} has adopted a mixed frame approach to solve the radiative transfer equation for moving fluid consistently.
After integrating the radiative transfer equation over frequency,  the equation becomes
\begin{eqnarray}
\frac{1}{c}\frac{\partial I}{\partial t}+ {\bf n}\cdot\nabla I= S(I,{\bf n})\,.
\end{eqnarray}
After carrying out the transport step in the lab frame, the source terms on the right hand side are added. But instead of
adding the source term $S(I,{\bf n})$ with all the $(v/c)$ and $(v/c)^2$ corrections to the intensity, 
the lab frame specific intensity $I({\bf n})$ at angle ${\bf n}$ is first transformed to the comoving
frame intensity $I_{0}({\bf n_0})$ via Lorentz transformation. Then the source terms in the comoving frame  ($S_{0}(I_{0},{\bf n_0})$)  are added to $I_{0}({\bf n_0})$,
\begin{eqnarray}
S_0(I_0,{\bf n_0})=\sigma_{a,R}\left(\frac{a_r T^4}{4\pi}-I_0\right)+\sigma_s \left(J_0-I_0\right)\nonumber\\
+\left(\sigma_{a,P}-\sigma_{a,R}\right)\left(\frac{a_rT^4}{4\pi}-J_0\right)\,,\label{eq:iequationnuint}
\end{eqnarray}
where $\sigma_{a,R}=\kappa_{a,R}\times\rho$, and $\sigma_{a,P}=\kappa_{a,P}\times\rho$. $\kappa_{a,R}$ and $\kappa_{a,P}$ are the Rosseland mean and Planck mean opacities. 
After this step to update $I_{0}({\bf n_0})$, $I_{0}({\bf n_0})$ are transformed back to the lab frame via Lorentz transformation. Then, 
the radiation momentum and energy source terms which are used in the fluid equations are calculated by the differences between the angular quadratures of
$I({\bf n})$ in the lab frame before and after adding the source terms.

For our particular FU Ori problem, we find that using the higher order PPM scheme \citep{Colella1984} for the transport step is crucial for deriving the correct radiation fields in the extremely optically thick regime (see Section 3.2). Thus, the PPM scheme has been used in all our simulations for solving the radiative transfer equation. 
Since the characteristic speed in the transport step is the speed of light, solving this equation explicitly requires very small numerical timesteps. Thus, we adopt the reduced speed of light approach
as in \cite{Zhang2018}. We reduce the speed of light by a factor of 1000, which still achieves a good timescale separation between the radiation transport and fluid dynamics. More discussions and tests on the reduced speed of light approach are in Section 3.2.
We solve the radiative transfer equation along 80 rays
in different directions. Integration of the specific intensity over angles yields various radiation quantities and source terms for the fluid equations. 

The opacity that is adopted in the radiative transfer equation is generated in \cite{Zhu2007,Zhu2009b}. 
With this opacity, \cite{Zhu2007} find an excellent agreement between the synthetic spectral energy distributions and observations for FU Ori.
This gives us great confidence to adopt it in this work for FU Ori MHD simulations.
Both Rosseland mean and Planck mean opacities are shown in Figure \ref{fig:rossplanck}.
The dust opacity that is below $\sim$1500 K is derived based on the prescription in
\cite{DAlessio2001}. 
The molecular, atomic, and ionized
gas opacities have been calculated using the Opacity Distribution
Function (ODF) method \citep{Sbordone2004, Castelli2004, Kurucz2005}
which is a statistical approach to handling line blanketing when
millions of lines are present in a small wavelength range
\citep{Kurucz1974}. More details on these opacities can be found in \cite{Zhu2009b} and \cite{KeithWardle2014}.
On the other hand, we adopt a simple equation of state with a constant $\gamma=5/3$ and $\mu=1$ to avoid any complications due
to the change of $\gamma$ and $\mu$ with the temperature. 

Our grid setup is similar to \cite{ZhuStone2018}, where the whole 4$\pi$ sphere is covered by the
spherical-polar ($r$, $\theta$, $\phi$)  grids with the special polar boundary condition in the $\theta$ direction (details in the appendix of \citealt{ZhuStone2018}). 
The grid is uniformly spaced in ln($r$), $\theta$, $\phi$ with 128$\times$64$\times$64 grid cells
in the domain of [ln(0.25), ln(100)]$\times$[$0$, $\pi$]$\times$[0, 2$\pi$] at the root level. Two levels of mesh refinement 
have been adopted at the disk midplane  with $\theta=[\pi/4,3\pi/8]$ and $[5\pi/8,3\pi/4]$ for the first level and $\theta=[3\pi/8,5\pi/8]$ for the second level. The outflow boundary conditions for flow variables, magnetic fields, and radiation fields have been adopted at both the inner and outer radial boundaries. Additionally, $v_{r}$ at the radial boundaries is set to prevent the inflow to the simulation domain.

The disk's initial density, temperature, and velocity profiles are also similar to \cite{ZhuStone2018} but with the midplane density slope of $p$=-2.125, the temperature slope of
$q=-3/4$, and $H/R$=0.2 at $R$=1. 
 This structure is consistent with the structure of a viscously heated $\alpha$ disk. The initial disk scale height is thus resolved by 16 grids with two levels of mesh refinement.
 The density floor is also similar to \cite{ZhuStone2018} except that an additional factor
of $r_{min}/r$ was multiplied to Equation (10) of \cite{ZhuStone2018} to further decrease the floor value at the disk atmosphere.

Simulations with both net vertical and net toroidal magnetic fields have been carried out. The net vertical field setup is similar to that in \cite{ZhuStone2018} with a constant plasma $\beta$ at the disk midplane initially.
In the net toroidal field simulations, magnetic fields are only present within 2 disk scale heights above and below the midplane initially, and the plasma $\beta$ is a constant anywhere within this region.

\subsection{FU Ori Parameters and Simulation Runs}
Our simulations adopt the disk parameters that are consistent with FU Ori observations. The detailed
disk atmospheric modeling \citep{Zhu2007, Zhu2008} suggests that FU Ori's inner accretion disk extends from 5 $R_{\odot}$ to $\sim$ 1 au with an accretion rate
of $2.4\times10^{-4}\msunyr$. The mass of the central star is 0.3 M$_{\odot}$. The rotational axis of the disk is 55$^{o}$ inclined with respect to our line of sight. 
Although these derived parameters are subject to change
due to the recent Gaia distance measurement and ALMA disk inclination measurement for FU Ori (see Section 5.3), we will use these numbers as a guidance for our simulation parameters. 

The length unit ($R=1$) in the simulation is chosen as 0.1 au so that the whole domain extends from 5 $R_{\odot}$ to 10 au. The density unit is chosen as
$10^{-8}$ g/cm$^3$ and the initial midplane density is $10^{-7}$ g/cm$^3$ at 0.1 au. The time unit is chosen as 1/$\Omega$ at 0.1 au around a 0.3 M$_{\odot}$ star.
In this paper, we use $T_{0}$ to represent the orbital period (2$\pi$/$\Omega$) at 0.1 au around a 0.3 $M_{\odot}$ star, which is 21 days.  With these units, the initial disk surface density is
  $\Sigma_0(r)= 7.5\times 10^{4} (R/0.1 {\rm au})^{-1} {g /cm^2}$.

Three main simulations have been carried out: 1) the disk that is initially threaded by net vertical magnetic fields with the strength of $\beta_{0}=1000$ at the disk midplane, labeled as V1000,
2) the disk that is threaded by vertical fields with $\beta_{0}=10^4$, labeled as V1e4, and 3) the disk that is initially threaded by net toroidal fields with the strength of $\beta_{0}$=100,
labeled as T100. We run these simulations to T$\sim$60 $T_{0}$, which is equivalent to $\sim$3 years. This time is equivalent to 500 innermost orbits in the simulation,
and the disk at $R=1$ has reached to a steady state as shown below. 

\subsection{Code Tests}

\begin{figure}
\includegraphics[trim=1mm 0.5mm 6mm 40mm, clip, width=3.3in]{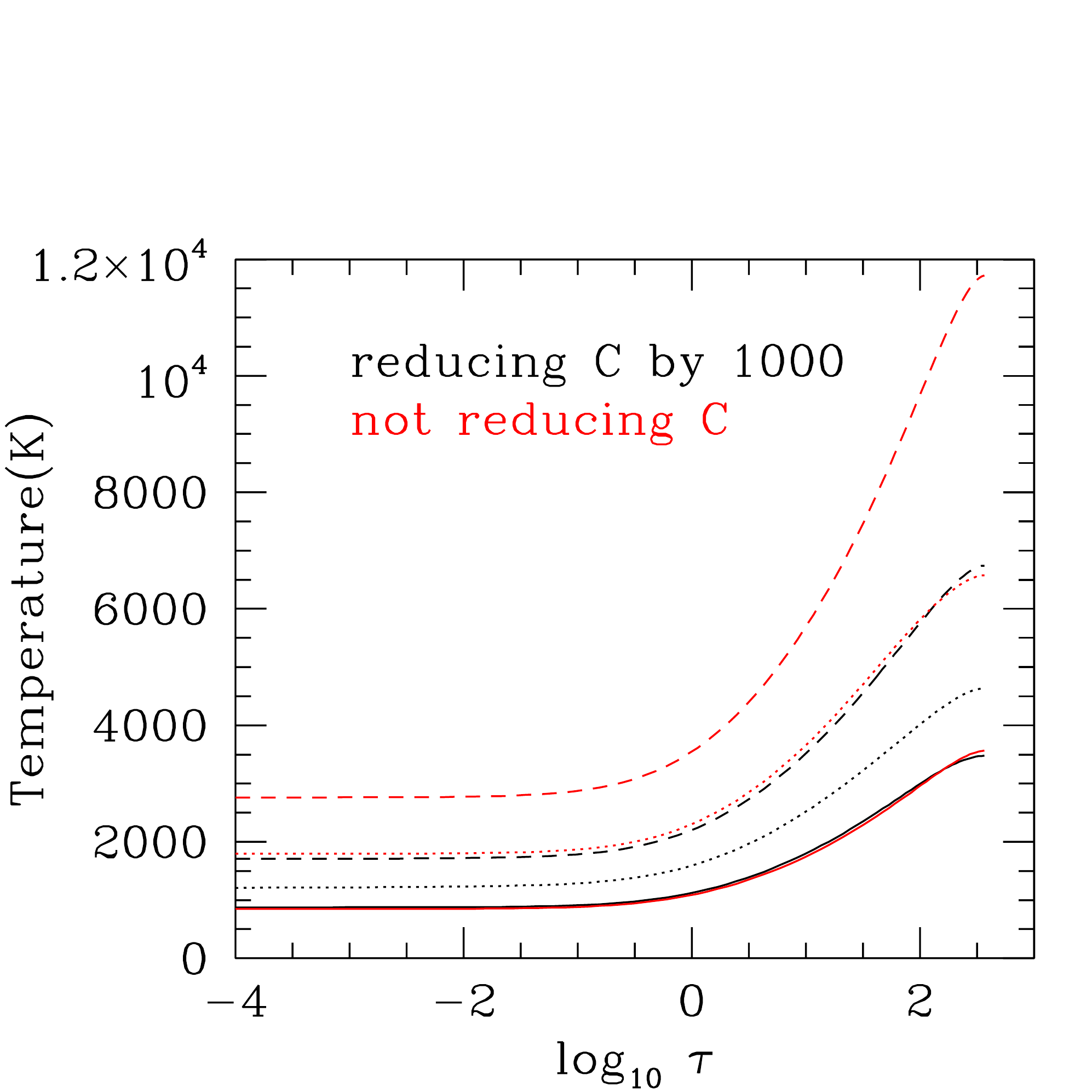}
\caption{Plane-parallel atmosphere tests similar to Figure \ref{fig:test} but with a sudden increase of the heating rate. 
With a normal heating rate, the disk reaches to a steady state after 2 $T_0$ (the solid black and red curves). Then, the heating rate suddenly jumps
to a value that is 100 times higher. After another 0.1 $T_0$, the disk thermal structures are shown as the dotted  curves. Then, after another 0.4 $T_0$, the disk thermal 
structures are shown as dashed curves. The adopted absorption opacity is 0.1 cm$^2$/g.
Clearly, using the reduced speed of light approach increases the timescale of radiation escaping the disk. 
 \label{fig:testred}}
\end{figure}

\begin{figure*}
\includegraphics[width=6.9in]{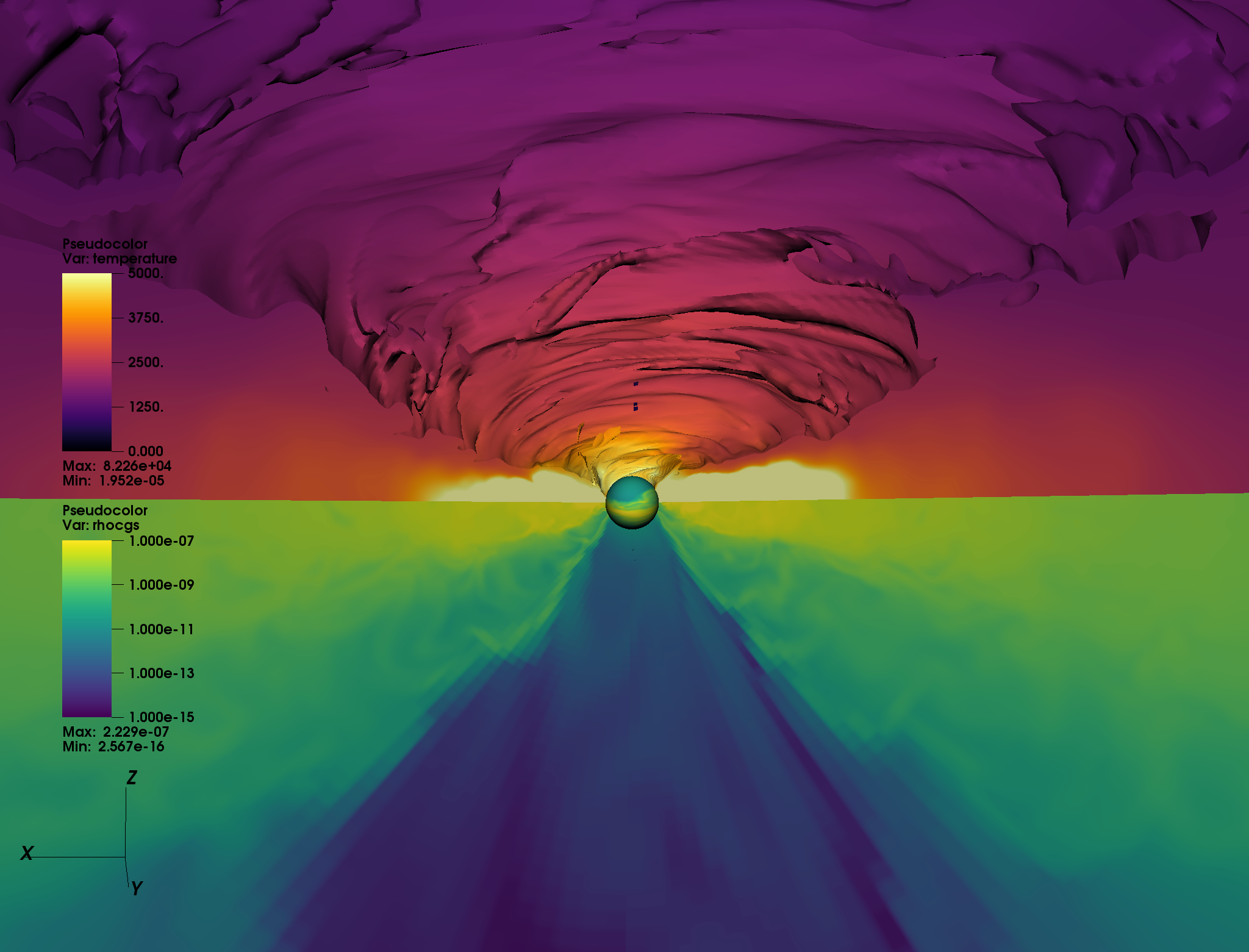}
\caption{The poloidal slice of the temperature (the upper half) and density (the lower half) from the V1000 case at 50 $T_{0}$. 
This illustrated region represents FU Ori disk within 0.5 au from the central star.
For the upper half of the image, the disk's photosphere is illustrated with the iso-surface having $\rho\kappa_{R}\times 0.1 au=10$.  
\label{fig:visit}}
\end{figure*}

\begin{figure*}
\includegraphics[trim=0mm 0mm 0mm 0mm, clip, width=6.6in]{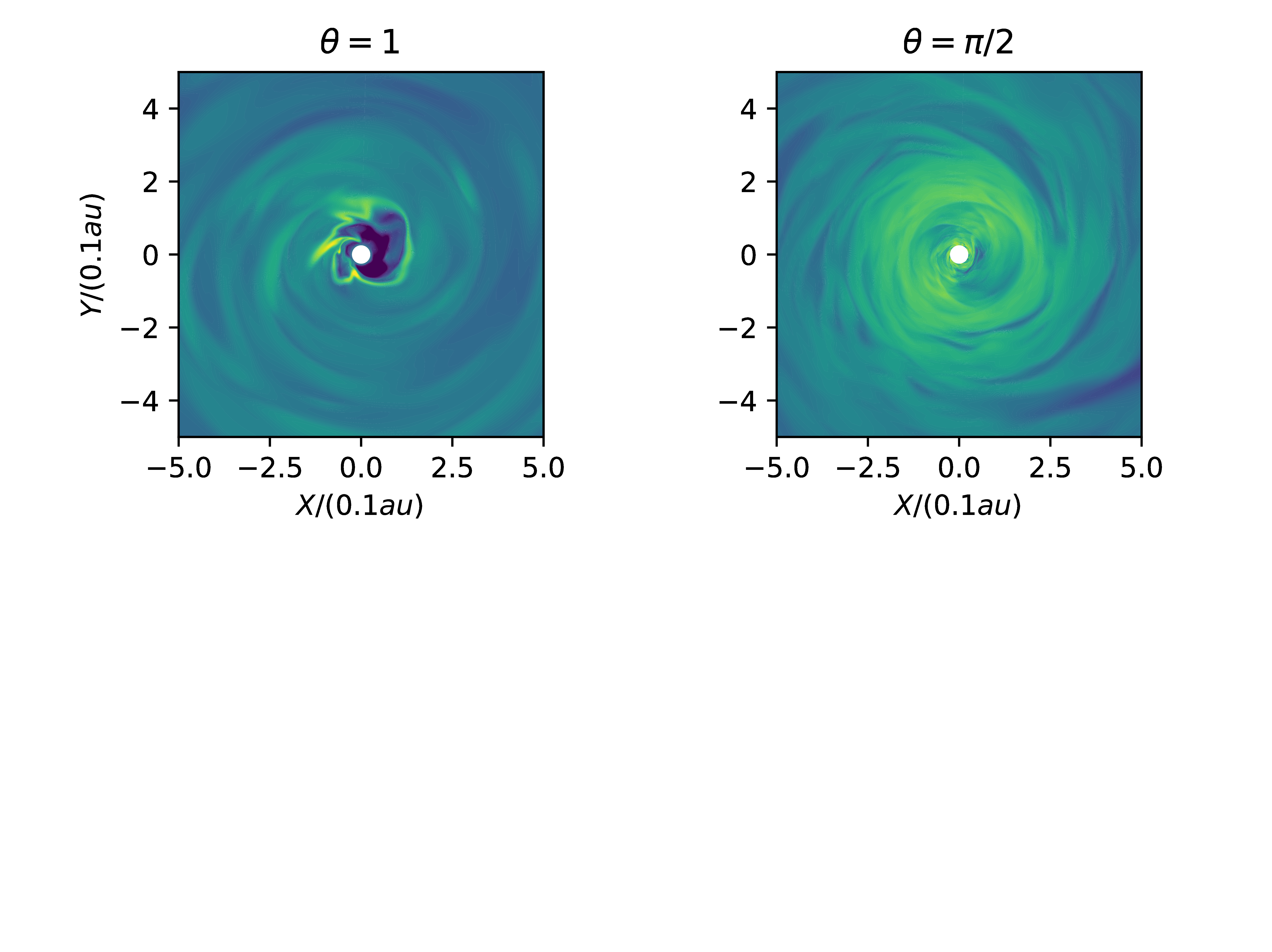}
\vspace{-57mm}
\caption{The contours of log$_{10}\rho$ at the $\theta=1$ plane (the left panel) and the midplane (the right panel) at  50 $T_{0}$. The color range
represents three orders of magnitude change of density in both panels.
 \label{fig:twodpolar}}
\end{figure*}

\begin{figure*}
\includegraphics[width=6.6in]{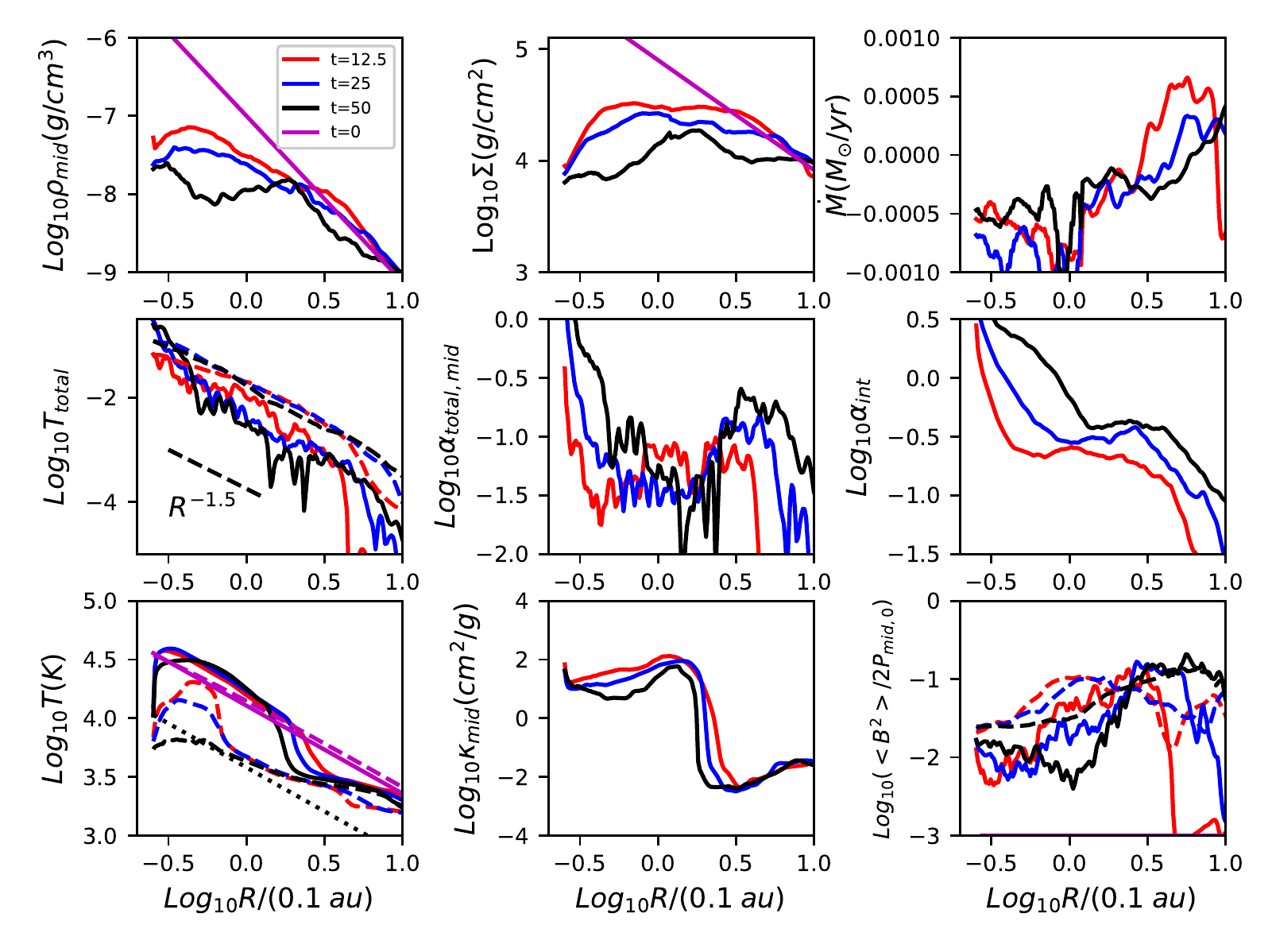}
\caption{The disk midplane density, surface density, mass accretion rate (upper  panels),
stresses (the solid curves are $r\phi$ stresses at the midplane while the dashed curves are the vertically integrated $R\phi$ stresses), midplane $\alpha$, vertically integrated $\alpha$ (middle panels),
 temperature, midplane Rosseland mean opacity,  and 
 $\langle B^2\rangle/2 P_{mid,0}$ (lower panels)  at different times. $\alpha_{total}$ and $\alpha_{int}$ are calculated with the $r\phi$ and $R\phi$ stresses respectively. In the temperature and  $\langle B^2\rangle/2 P_{mid,0}$ (where $P_{mid,0}$ is the midplane pressure from the initial condition) panels,  the solid curves are the midplane quantities and the dashed curves are the quantities along $r$ at $\theta=0.78$ (where the photosphere roughly sits). The black dotted line 
in the temperature panel is from Equation \ref{eq:teffsim} with an accretion rate of 4$\times$10$^{-4}\msunyr$ around a 0.3 $M_{\odot}$ star. \label{fig:onedradialfig2rad125}}
\end{figure*}

\begin{figure*}
\includegraphics[width=6.6in]{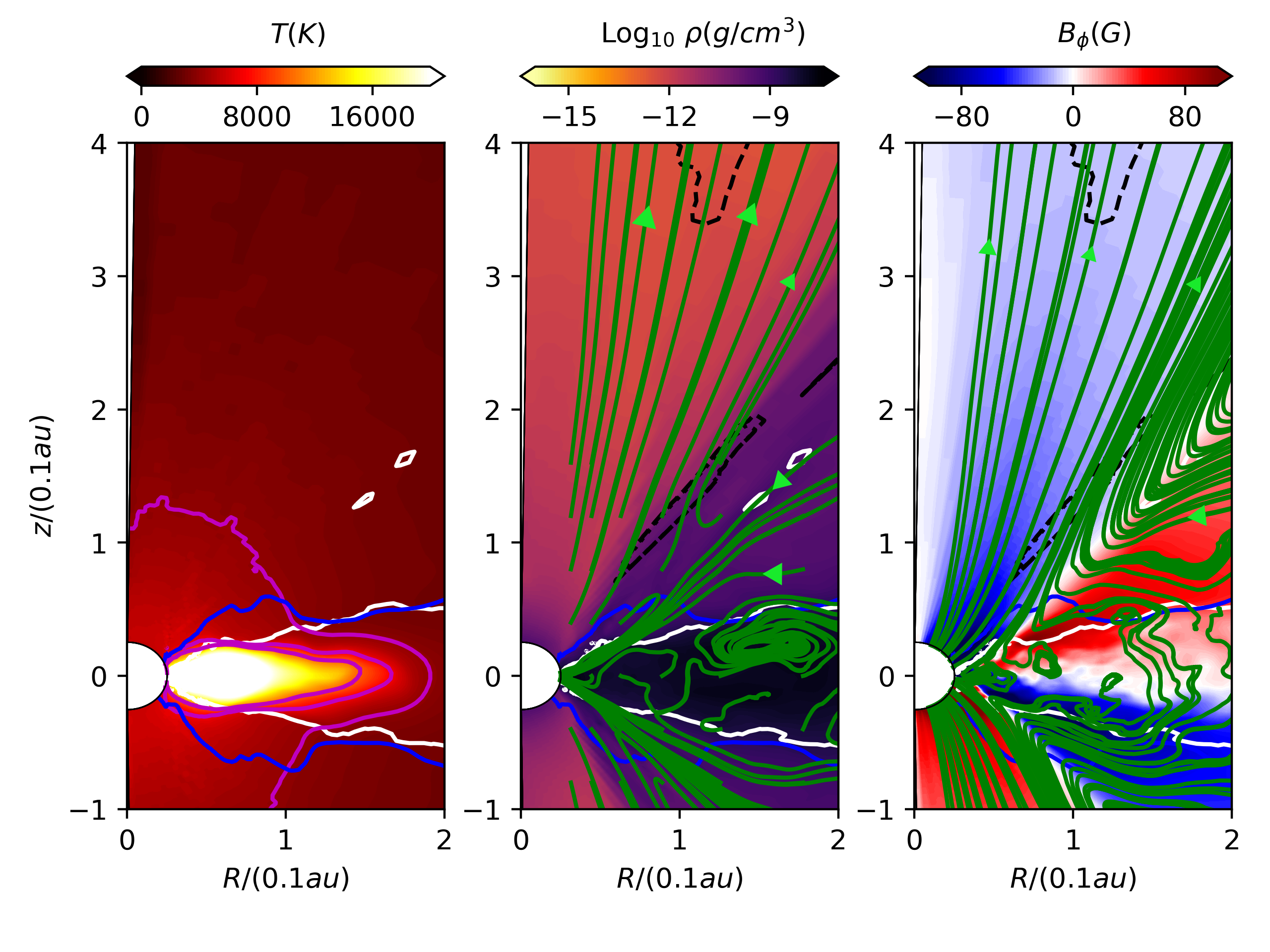}
\caption{The azimuthally averaged temperature (the left panel), density (the middle panel), and $B_{\phi}$ (the right panel) for the V1000 case at 50 $T_{0}$.
The green lines in the middle panel are the streamlines for the poloidal velocity fields, while the green lines in the right panel are the streamlines for the poloidal magnetic fields (the direction of
the magnetic fields
at the upper boundary is pointing upwards). 
The white contours in all these panels are the $\beta=1$ surfaces. The purple curves in the left panel are the contours where $T$=4000, 7000, and 10000 K.  The blue
curves in the three panels are the $\tau_{R}=1$ surfaces. The dashed curves in the middle and right panels are the Alfven surfaces. 
\label{fig:twodrhoBstrong}}
\end{figure*}

\begin{figure*}
\includegraphics[trim=0mm 3.5mm 0mm 13mm, clip, width=5.5in]{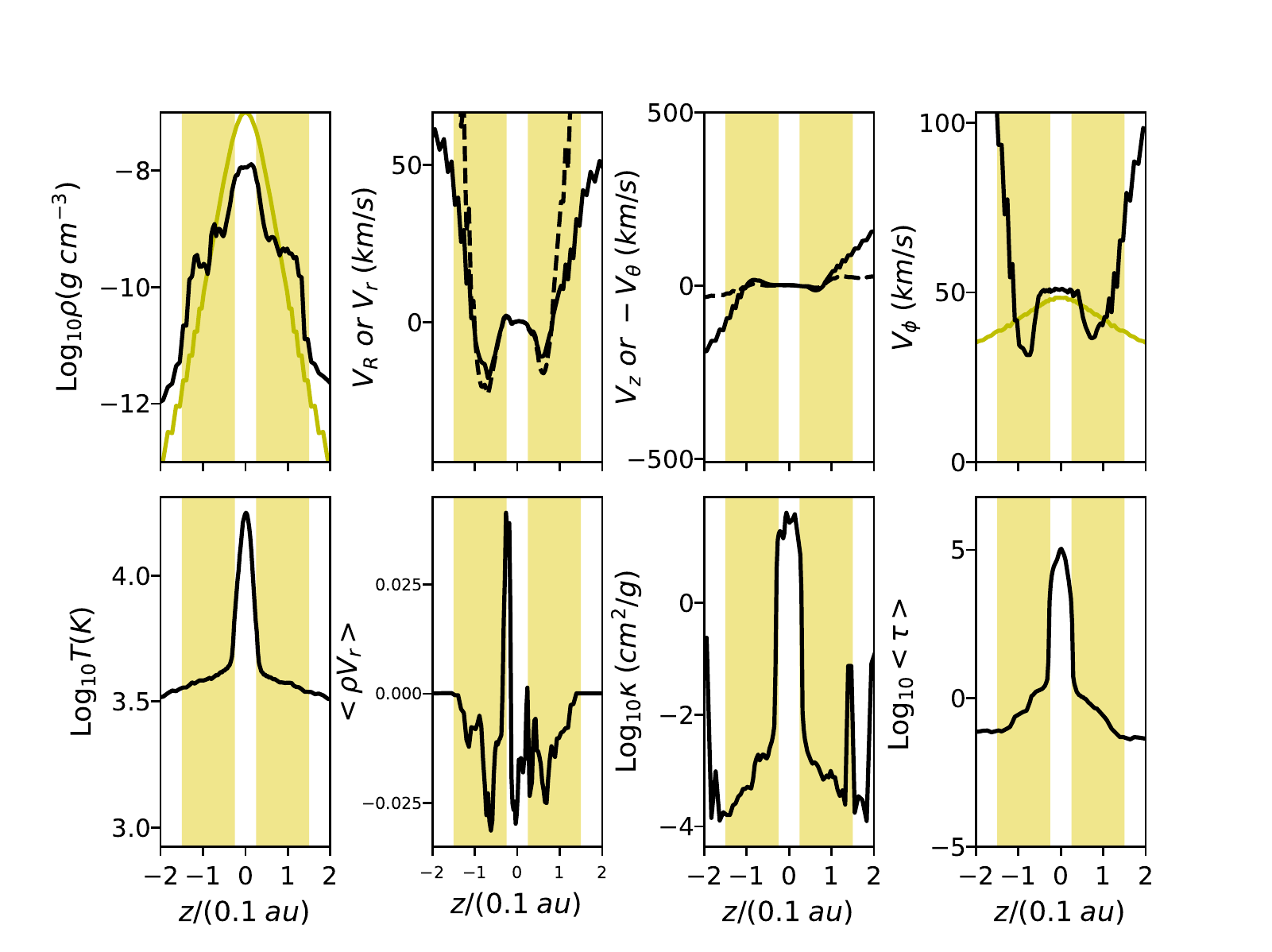}
\caption{Density, velocities, temperature, mass flux, opacity, and optical depth along the $z$ direction at $R=$ 0.1 au at 50 $T_{0}$. 
The quantities have been averaged azimuthally. The dashed curves in the velocity panels show the velocity components in the spherical-polar
coordinates ($V_{r}$ and $-V_{\theta}$). The yellow curves are from the initial condition. The yellow shaded region labels the surface accreting region. 
Note the fast inward flow at the disk surface. \label{fig:onedrfig3rad500v2strongpaper}}
\end{figure*}

\begin{figure*}
\includegraphics[trim=0mm 3.5mm 0mm 13mm, clip, width=5.5in]{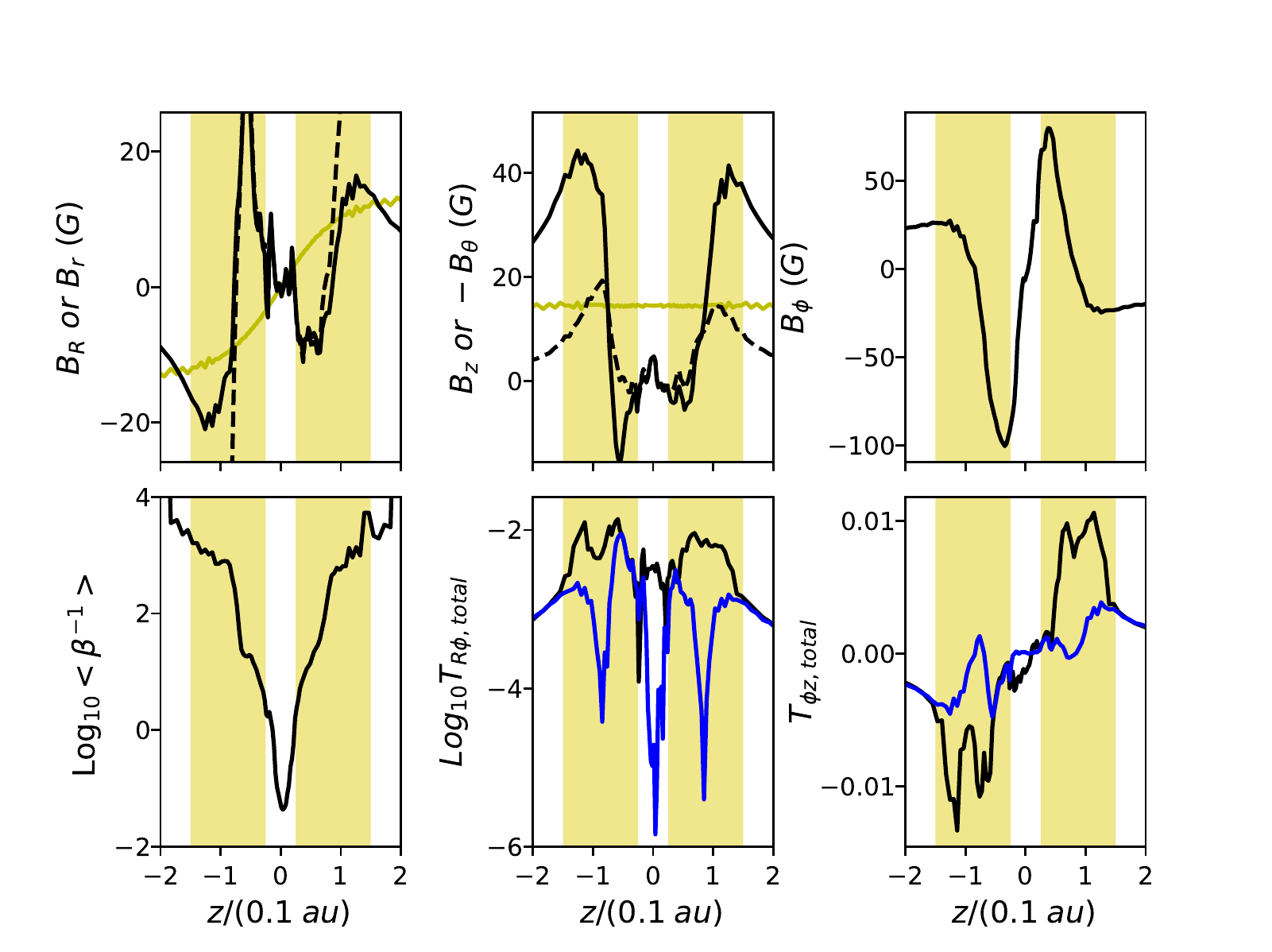}
\caption{Similar to Figure \ref{fig:onedrfig3rad500v2strongpaper} but for quantities that are related to magnetic fields. The dashed curves are B components in the spherical-polar coordinates ($B_{r}$ and $-B_{\theta}$). The blue curves
in the $T_{R\phi}$ and $T_{\phi z}$ panels are magnetic stresses that are calculated using the mean fields, -$\overline{B_{R}}\times\overline{B_{\phi}}$ and -$ \overline{B_{z}}\times\overline{B_{\phi}}$. The mean fields are azimuthally averaged before being used to calculate the stress. \label{fig:onedrfig3rad500v2strongpaperb}}
\end{figure*}

\begin{figure}
\includegraphics[trim=0mm 0.5mm 0mm 0mm, clip, width=5.5in]{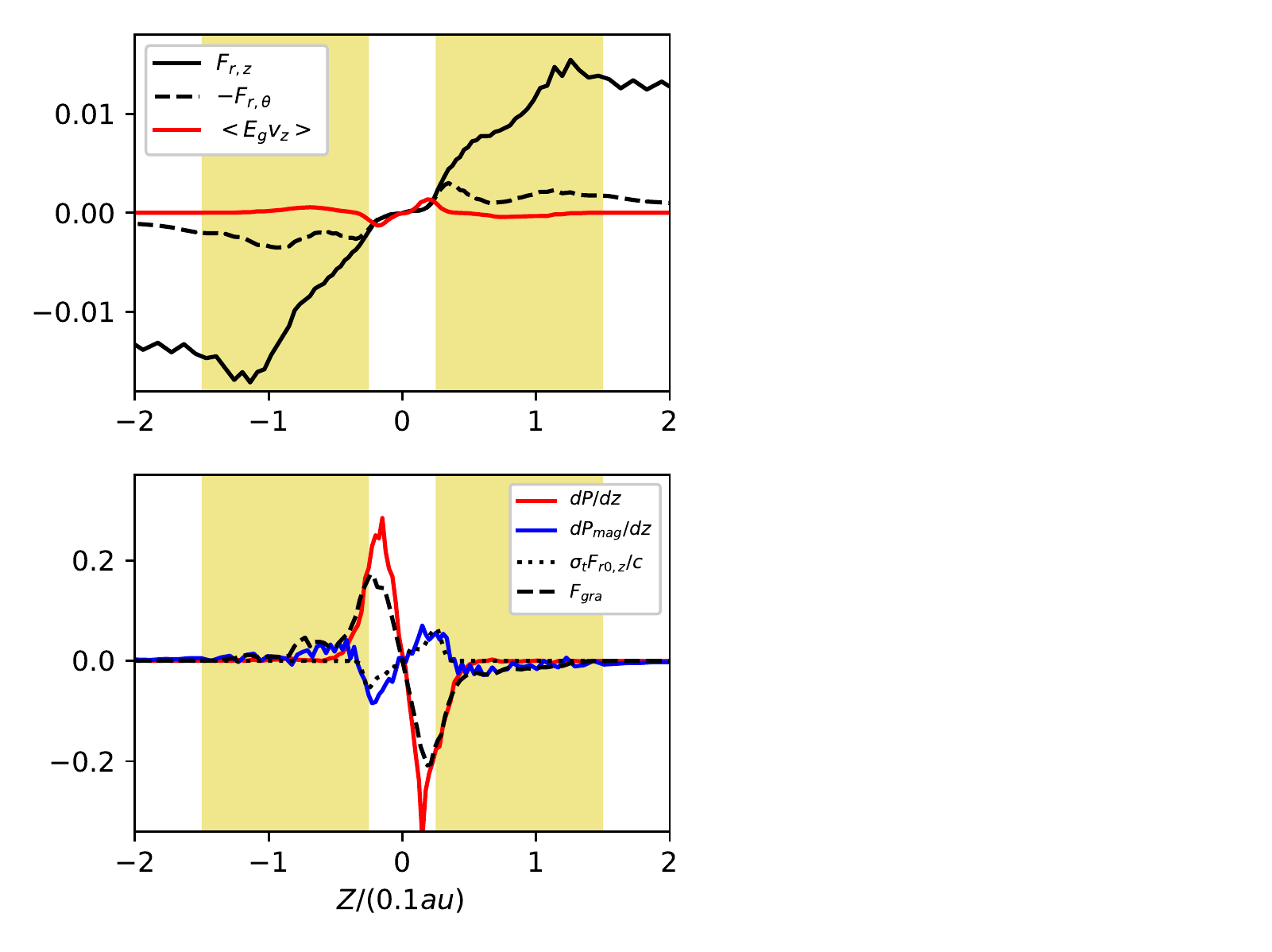}
\caption{The upper panel: the vertical energy flux at $R=0.1$ au due to radiation ($F_{r,z}$) and convection ($<E_{g}v_{z}>$). The bottom panel: the force balance in the vertical direction, including
the gravitational force ($F_{gra}$), the radiation force ($\sigma_t F_{r0,z}/c$), the gas pressure gradient ($dP/dz$), and the magnetic pressure gradient ($dP_{mag}/dz$). All quantities are averaged over both the azimuthal direction and time (45 to 50 $T_{0}$ with a 0.1 $T_{0}$ interval).  \label{fig:balance}}
\end{figure}

\begin{figure}
\includegraphics[trim=5mm 4.5mm 73mm 1mm, clip, width=3.in]{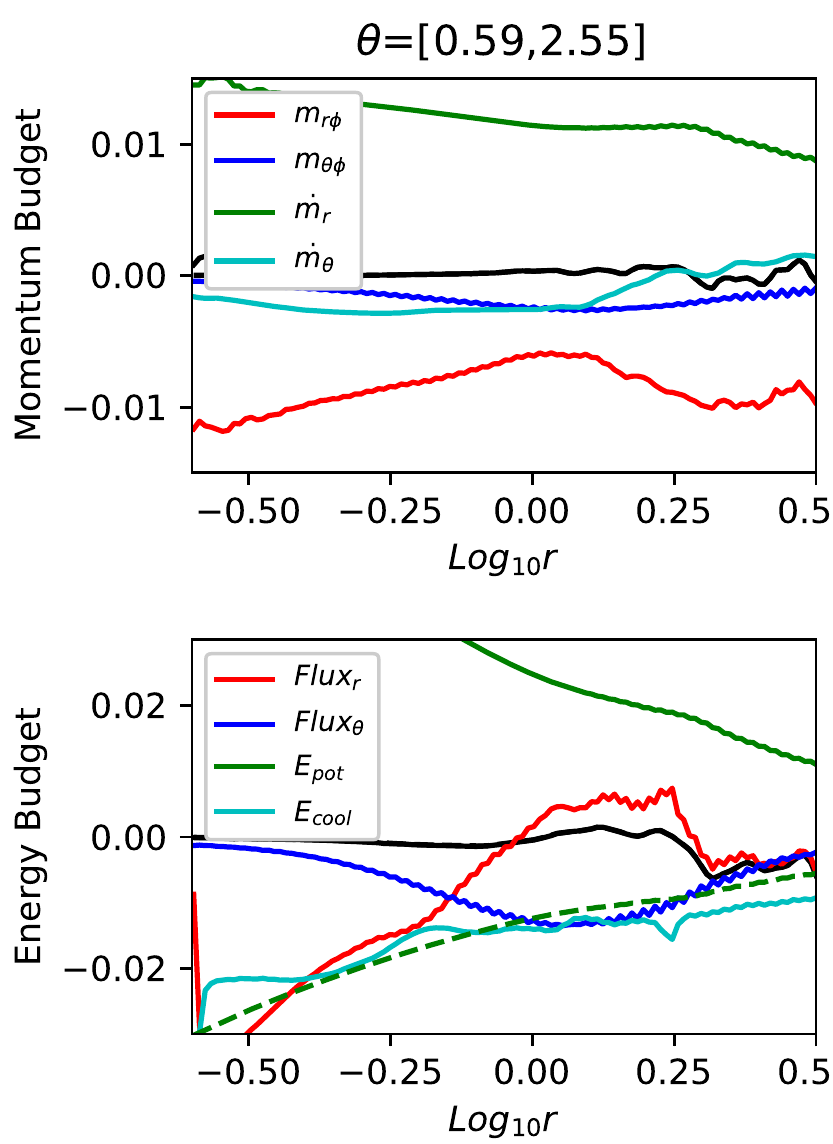}
\caption{The angular momentum (the upper panel) and energy (the lower panel) budget for our fiducial run (V1000). Various components
of the budget have been averaged over time (from $t=42 T_0$ to $52 T_0$) and integrated over space ($\theta$ from 0.59 to 2.55 to include both the accreting surface and the midplane region). The averaged
quantities have also been multiplied by $r^{3.5}$ so that these quantities are almost flat in radii. The green dashed curve in the lower panel is $-E_{pot}/2$ for comparison. The black curve in each panel is the addition of all the four components.
\label{fig:onedconfig13full2sin43w0p98papervert}}
\end{figure}

\begin{figure}
\includegraphics[trim=9mm 13.5mm 65mm 40mm, clip, width=3.in]{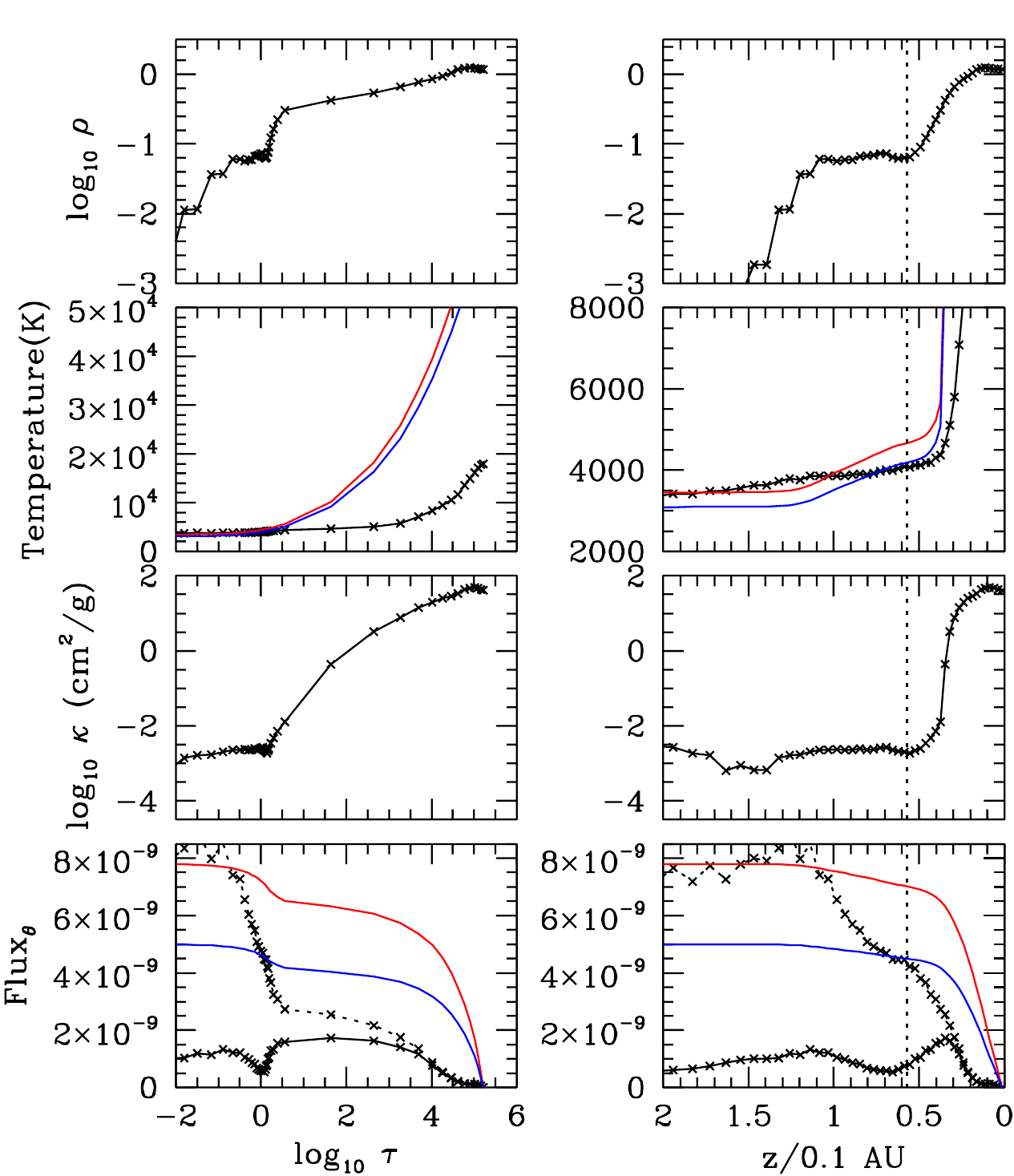}
\caption{The disk vertical structure along $R=0.1$ au with respect to $\tau$ starting from the  disk surface (left panels) or $z$ starting from the midplane (right panels).  
In the bottom panels, the crosses with the solid black curves are $F_{r,\theta}$, while the crosses with the dashed black curves are $F_{r,z}$. All quantities are averaged over both the azimuthal
direction and time (45 to 50 $T_0$ with a 0.1 $T_{0}$ interval).
The red and blue curves in the temperature and $F$ panels are the analytical solutions using Equation \ref{eq:tauana} with two different fluxes.
The red one uses the flux that is calculated with Equation \ref{eq:Qcoolteff} and the measured $\dot{M}$; the blue one uses the flux that is calculated with
Equation \ref{eq:teffsim} and the measured $\dot{M}$.  The black dotted line labels where $\tau_R=1$ in simulations.
\label{fig:Tstru}}
\end{figure}

\begin{figure*}
\includegraphics[width=6.6in]{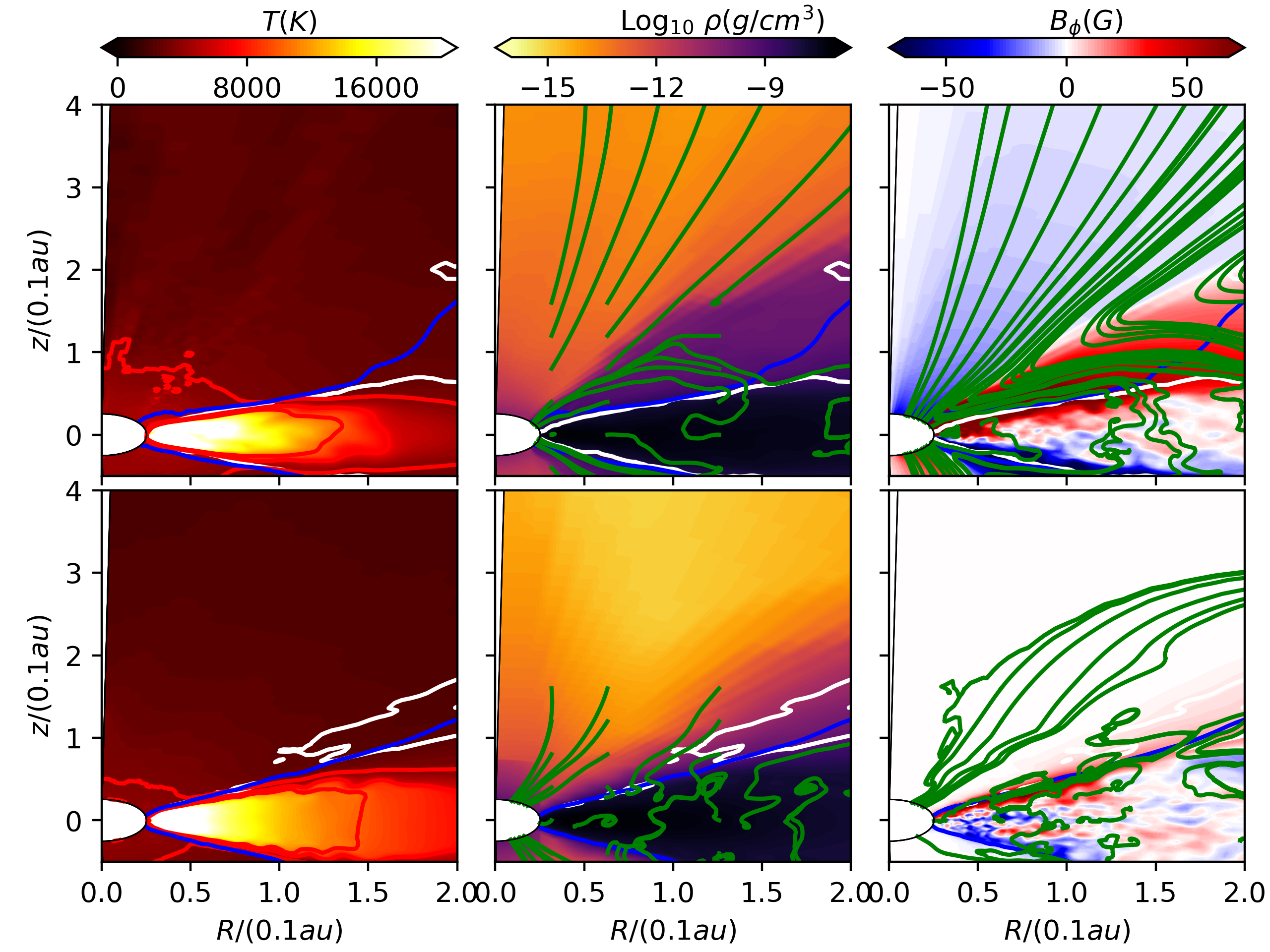}
\caption{Similar to Figure \ref{fig:twodrhoBstrong} but for the V1e4 case at $t=55 T_{0}$ (upper panels) and the T100 case at $t=60 T_0$ (lower panels). \label{fig:twodrhovelB600}}
\end{figure*}

\begin{figure*}
\includegraphics[trim=0mm 4mm 0mm 2mm, clip, width=5.in]{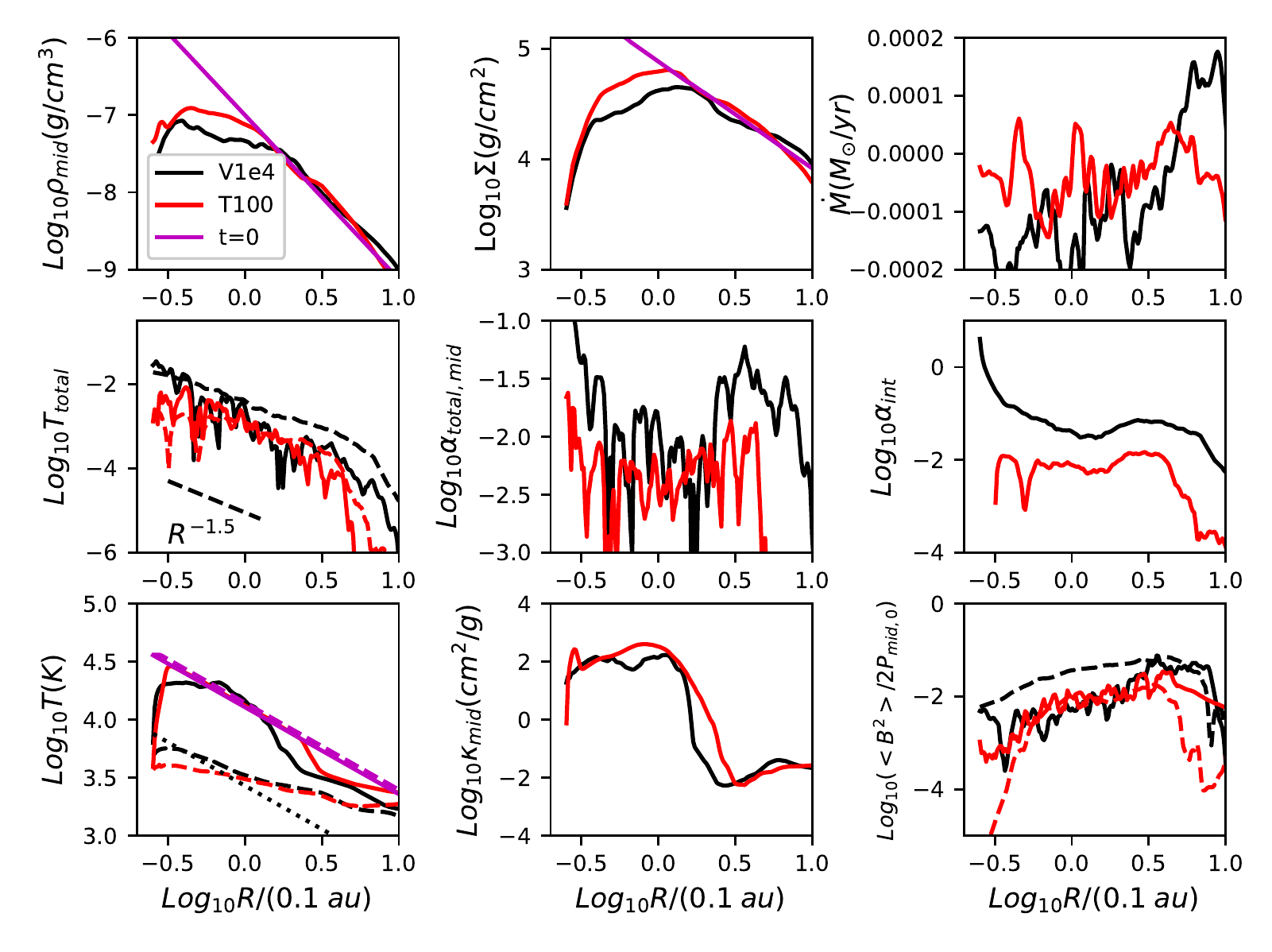}
\caption{Similar to Figure \ref{fig:onedradialfig2rad125} but for the V1e4 case at $t=55 T_{0}$ (black curves) and the T100 case at $t=60 T_0$ (red curves). In the  temperature and  $\langle B^2\rangle/2 P_{mid,0}$ panels,  the solid curves are the midplane quantities and the dashed curves are the quantities along $r$ at $\theta=1.1$  where the photosphere is.  The black dotted line 
in the temperature panel is from Equation \ref{eq:teffsim} with an accretion rate of 10$^{-4}\msunyr$. \label{fig:onedradialfig2rad550compare}}
\end{figure*}

\begin{figure*}
\includegraphics[trim=0mm 3.5mm 0mm 13mm, clip, width=5.5in]{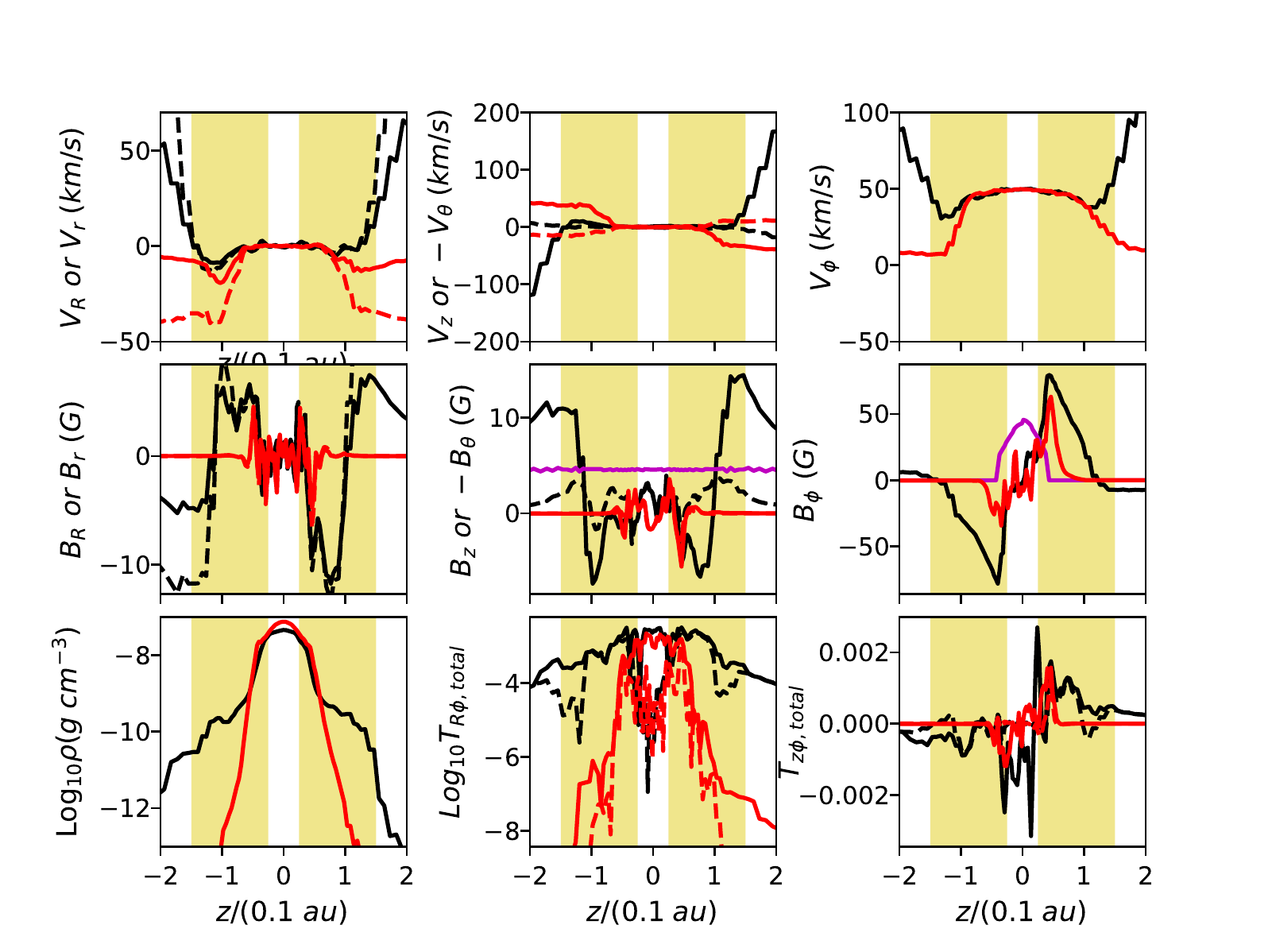}
\caption{Similar to Figures \ref{fig:onedrfig3rad500v2strongpaper} and \ref{fig:onedrfig3rad500v2strongpaperb} but for the V1e4 case (black curves) and  the T100 case (red curves). \label{fig:onedrfig3rad600v2strongcomparepaperb}}
\end{figure*}

Although the radiative transfer scheme has been tested extensively (e.g. \citealt{Jiang2014, Jiang2019a}), we still need to test
if the scheme is applicable to our particular FU Ori disk setup. 
Thus, we set up a 1-D plane-parallel atmosphere with a density profile of
\begin{equation}
\rho=\rho_{0} e^{-z^2/2H^2}\,,\label{eq:simpleatmosphere1}\\
\end{equation}
to represent the disk's vertical density structure at $R$=1 in our 3-D FU Ori simulations. $H$ is chosen as 0.02 au, and
$\rho_{0}$ is chosen as 10$^{-8}$ g/cm$^3$. All other parameters are the same as our 3-D FU Ori simulations.
 To maintain this density structure, we don't update the density and velocity during the simulation, and only allow the disk temperature to change. Only two rays have been used in 
 this setup so that we can use two-stream approximation to calculate the analytical solution. 

To represent the viscous heating in the accretion disk, we manually include a heating source term with the heating rate that is proportional to the disk local density ($\rho$) as
\begin{equation}
\frac{dE}{dt}=C\times \rho\,.
\end{equation}
We have done three tests, two of which are steady state tests with a constant $C$ and one of which is the increasing heat test where $C$ suddenly increases at some time.

In the steady state tests,
two different values of $C$ (0.0002316 and 0.02316 in the code unit) have been used to test if the disk can reach to the correct temperature in both low and high temperature regimes. 
The lower heating rate only heats the disk to $T\sim$ 10$^3$ K, when the opacity is dominated by the
dust and molecular opacities (the upper panels in Figure \ref{fig:test}). 
The higher heating rate heats the disk to $T\sim$ 10$^4$ K, when the opacity is
dominated by the free-free and bound-free opacities (the lower panels in Figure \ref{fig:test}).  

These steady state tests show that we can accurately simulate the disk thermal structure for some cases, but also reveal the limitation of our setup.
The black crosses in Figure \ref{fig:test} are results from simulations with 160 grids from -0.1 to 0.1 au (the same resolution as our 3-D simulations), 
while the red curves are from simulations with 1600 grids in the same domain range.
The blue curves in the middle panel are the analytical solutions of this problem solved with the two-stream approximation:
\begin{equation}
T(\tau)^4=\frac{3}{4}T_{eff}^4\left(\tau\left(1-\frac{\tau}{\tau_{tot}}\right)+\sqrt{\frac{1}{3}}\right)\,,\label{eq:tauana}
\end{equation}
where $\sigma T_{eff}^4$ is the flux emerging from one side of the disk and $\tau_{tot}$ is the total optical depth from both sides of the disk. 
Clearly, when the opacity is low (e.g. the upper panels), the simulations with different resolutions agree with the analytical solution very well, even if
the opacity has sharp changes among grids. On the other hand, when the opacity is high (e.g. the bottom panels), the optical depth can jump more than one order of magnitude
from one grid to another grid. As expected, this jump leads to large errors in the calculations. Unfortunately, even with the resolution that is 10 times higher (red curves in the lower panels), we still cannot recover the analytical solution accurately. 
One way to overcome this problem in future is using adaptive mesh-refinement for those grid cells having high optical depths. Overall, this test shows that, with our current setup, we may underestimate
the temperature of some extremely optically thick grid cells by a factor of 2.

Since FU Ori's disk temperature can change dramatically before and during the outburst, 
we also need to test if the code can capture the time evolution of the disk's temperature accurately. 
Especially, our adoption of the reduced speed of light approach may delay the escape of the radiation energy. This
is a particular concern when the disk is very optically thick \citep{SkinnerOstriker2013} since the diffusion timescale $L\tau/c$ can now be longer than the dynamical timescale. 
For a typical size scale of 0.1 au and an optical depth of 1000, the radiation diffusion timescale is $\sim$1 day.  Naively, we would think that decreasing the speed of light by 1000 will
increase the diffusion timescale to 1000 days, which is even longer than the total simulation timescale.  On the other hand, it can be shown that the formulation in \cite{Zhang2018}
guarantees that the radiative diffusion flux is the correct flux when the thermal energy of the gas dominates over the radiation energy. Thus, we should expect a correct diffusion timescale for 
our setup where the thermal energy of the gas always dominates.
However, one could also argue that 
the optically thick region is joined by the optically thin region, and the escape of the total energy will be controlled by the optically thin region so that
the disk will still cool/heat slower with the reduced speed of light approach. 

To resolve these concerns, we carry out a test with a suddenly increased heating rate. 
We fix the absorption opacity to be 0.1 cm$^2$/g in this test. Initially,
the disk is heated at the heating rate of $C$=0.0002316   for a period of 2 T$_0$ so that the disk reaches to a steady state. 
Then, we suddenly increase the heating rate by a factor of 100 and watch the subsequent disk evolution. 
As shown in Figure \ref{fig:testred}, the reduced speed of light approach indeed slows down the heating of the disk. 
On the other hand, the temperature structure at 0.5 $T_{0}$
after the heating event for the disk using the reduced speed of light approach (the black dashed curve) overlaps with the temperature structure
at 0.1 $T_0$ after the heating event for the disk using the normal speed of light (the red dotted curve).
Thus, the reduced speed of light approach increases the diffusion timescale by a factor of $\sim$ 5. This is larger than 1, but it is also much smaller than 1000 so that
the diffusion timescale is still much shorter than the simulation timescale. 
Nevertheless, since the reduced speed of light approach increases the diffusion timescale to $\sim T_0$, we cannot trust short timescale variations of the radiation field in the simulations,
and we can only study the state when the disk is relatively steady for  the orbital timescale. Thus, in this paper, we only focus on the disk at the steady state with a constant accretion rate instead of
discussing the outburst stage when the disk suddenly brightens by orders of magnitude within a short period of time. 

\section{Results}
The temperature and density structures of our fiducial model (V1000) at 50 $T_0$ are shown in Figure \ref{fig:visit}.
We can see that the disk atmosphere at $z\sim R$ still has a significant density, which is similar to the disk structure
in \cite{ZhuStone2018}.  With the radiative transfer included in our simulations, we can now study the disk's temperature structure. 
The disk's temperature is quite high ($\gtrsim$5000 K) close to the central star ($\lesssim$0.15 au). There is a sharp temperature jump
around 0.15 au, indicating that the inner disk is at the upper branch of the equilibrium ``S" curve which is dominated by the bound-free and free-free opacity while the outer
disk is at the lower branch of the equilibrium ``S'' curve ($\lesssim$2500 K) which is dominated by the molecular opacity. We also use
 $\rho \kappa_{R}\times$0.1 au $\sim$10 to illustrate the disk's photosphere. Clearly, the photosphere is hotter at the inner disk than at
 the outer disk, and the photosphere is not smooth having noticeable structures. 
 Figure \ref{fig:twodpolar} shows the density structure at the disk surface and the midplane. At the midplane, we clearly see spiral arms similar to those found in \cite{Mishra2019}.
 On the other hand, the disk surface has filamentary structure due to surface accretion, as found in \cite{ZhuStone2018,Suriano2018}.
 Due to these large scale structures at the photosphere, we expect
 that FU Ori has short timescale variations which have been implied by observations \citep{Kenyon2000,Herbig2003,Powell2012,Siwak2013}. 

After running for 50 $T_0$, our fiducial model has reached to a steady state within $R\sim$0.5 au, i.e., the inner factor of $\sim$ 20 in radius, as evident in
Figure \ref{fig:onedradialfig2rad125}.
From the mass accretion rate panel (the upper right panel), we can see that, the region that is accreting inwards expands with time since
the outer disk region takes more time for MRI to grow. At 50 $T_{0}$, the region within 0.5 au, i.e., the inner factor of $\sim$20 in radius, 
accretes inwards at a steady rate. Such constant accretion rates are also consistent with the stress profiles shown in the middle left panel. 
The vertically integrated $R\phi$ stress follows $R^{-1.5}$ and this leads to a constant accretion rate based on Equation \ref{eq:stresscyl}.
Such accretion and stress structures are very similar to the global MHD simulations with the locally isothermal equation of state (compared with Figure 3 in \citealt{ZhuStone2018}).

However, other quantities shown in Figure \ref{fig:onedradialfig2rad125} are drastically different from those in Figure 3 of \cite{ZhuStone2018}.
For example, the surface density in Figure \ref{fig:onedradialfig2rad125} is almost flat, which is different from $R^{-0.6}$ in \cite{ZhuStone2018}.
The midplane $\alpha$ is also flat compared with $R^{0.5}$ in \cite{ZhuStone2018}. Such differences are likely due to the temperature structure at the
midplane. In the viscous heating dominated disk presented here, the midplane temperature follows $\sim R^{-3/4}$ (the lower left panel), 
while, in the locally isothermal simulations, the midplane temperature
follows $R^{-1/2}$. Another evidence that the midplane temperature affects the $\alpha$ profile is that, at $R\sim0.15$AU where the midplane
temperature jumps down, the $\alpha_{total,mid}$ there jumps up so that the total stress $T_{total}$ is still smooth. It is quite surprising
that the accretion and stress profiles are smooth despite the jump of disk temperature. Considering that most stress is from the magnetic stress, 
this implies that the disk's accretion structure is mainly
controlled by the global geometry of magnetic fields and is insensitive to the disk local temperature.  The magnetic fields at the midplane
and $\theta=0.78$ are shown in the lower right panel, and we can see that the field strength changes smoothly in the disk despite the temperature jump at $R\sim$0.15 au.

\subsection{Accretion Structure}
The flow structure in MHD disks is tightly coupled with the magnetic field geometry. 
Magnetic fields determine the accretion structure while the accretion process drags and alters the magnetic fields.
We plot the azimuthally averaged temperature, density, and magnetic field structures for our fiducial run in Figure \ref{fig:twodrhoBstrong}.

The velocity and magnetic field structures are remarkably similar to the ``surface accretion'' picture in locally isothermal disks with net vertical fields \citep{ZhuStone2018}.
Although we called such surface accretion as "coronal accretion" in \cite{ZhuStone2018} following \cite{Beckwith2009}, the accreting surface may not be as hot as Sun's ``corona'' that exceeds 10$^6$ Kelvin (as shown in this work and \citealt{Jiang2019b}).  On the other hand, the accreting surface is more associated with the strong magnetic fields ($\beta\lesssim$1, or called magnetically elevated in \citealt{Mishra2019}).
Thus, in this work, we call this structure as ``surface accretion'' instead.
The flow structure can be separated into three regions from the midplane upwards: the disk region which is dominated by MRI turbulence, the surface accreting region which is 
above the $\beta=1$ surface and extends all the way to  $z\sim R$, and the disk wind region (with $v_{r}>0$) at $z\gtrsim R$.
The accretion flow mainly occurs at the surface, as shown in the middle panel of
Figure \ref{fig:twodrhoBstrong} where the velocity streamlines are towards the star in the surface accreting region. 
Such surface inflow drags magnetic fields inwards so that the fields are pinched at the disk surface (the right panel of
Figure  \ref{fig:twodrhoBstrong}). Due to the increase of the Keplerian rotation speed towards the inner disk,
these dragged-in magnetic fields are sheared azimuthally, leading to fields with opposite $B_{\phi}$ between the lower and higher surface regions. 
Such surface accretion has been seen as early as \cite{StoneNorman1994} and
recently  in several simulations
\citep{Beckwith2009, SuzukiInutsuka2009, ZhuStone2018, Suriano2018, Takasao2018, Mishra2019, Jiang2019b}.
Analytical works by \cite{GuiletOgilvie2012, GuiletOgilvie2013} have also seen such surface accretion
when the turbulent viscosity and diffusivity are considered in their analytical works.

On the other hand, our radiation MHD simulations reveal new information on the disk thermal structure, especially the position of the disk photosphere.
The left panel of Figure  \ref{fig:twodrhoBstrong} shows that the thermal radiation field is very smooth except at the sharp jump $\sim$ 0.15 au
separating the two states that reside at the upper and lower branches of the ``S'' curve. If we integrate the Rosseland mean opacity along 
the $z$ direction (starting from 20$^o$ off the axis
to avoid the coarse grids at the pole), the derived $\tau_{R}=1$ surface is plotted as the blue curves in all three panels. We can see that
the $\tau_R=1$ surface is at the wind base or upper surface accreting region at the inner disk ($\lesssim$0.07 au) and within the lower surface accreting region at the outer disk ($\gtrsim$0.07 au). 
Thus, $B_{\phi}$ derived from the atomic lines at the photosphere could have opposite directions depending on where these lines are produced. 
This has important implications for the B field measurements of FU Ori, which will be discussed in greater detail in Section 5.1. 
This transition radius $\sim$ 0.07 au, which roughly corresponds to the filamentary structure shown in Figure \ref{fig:twodpolar}, may also be related to the periodic variability at 10-15 days found in \cite{Herbig2003,Powell2012,Siwak2013,Siwak2018}. 

To understand the disk's accretion structure quantitatively, we plot the vertical profiles of various quantities at 0.1 au in Figure \ref{fig:onedrfig3rad500v2strongpaper}.
The yellow shaded region is the surface accreting region. We see that the density flattens out in the surface accreting region, and the radial accretion velocity can reach 20 km/s there (the $v_{R}$ panel). Considering
that the Keplerian velocity is 50 km/s at 0.1 au, the surface inflow velocity is $\sim$40\% of the Keplerian velocity. Due to the high speed, most disk mass is accreted through this surface accreting region despite its low density (the $\rho v_{r}$ panel). The azimuthal velocity also deviates from the Keplerian velocity. In the surface accreting region,
the lowest azimuthal velocity can reach to $60\%$ of the Keplerian velocity (the $v_{\phi}$ panel). Such low azimuthal velocity and high radial velocity can be understood 
as magnetic breaking by the midplane so that the surface loses angular momentum and falls inwards. 
The midplane is very hot with a high opacity. Here at $R$=0.1 au, the disk's photosphere ($\tau_R=1$) is within the surface accreting region  (the $\tau$ panel). 

The magnetic field structure at $R$=0.1 au is shown in Figure \ref{fig:onedrfig3rad500v2strongpaperb}. 
The surface inflow drags the initial vertical magnetic fields inwards, pinching the magnetic fields at the disk surface.
The radial component of the magnetic fields in the surface accreting region has been sheared by the Keplerian rotation to produce a strong azimuthal component. The azimuthal $B$ component can reach to 100 G, which is $\sim$5 times the radial $B$ component. The combination of $B_{z}$ and $B_{\phi}$ produces
positive $\partial T_{\phi z}/\partial z$ at the base of the surface accreting region. Using Equation \ref{eq:stresscyl}, we can see that this $T_{\phi z}$ leads to the inward accretion of the surface. In other words, the midplane is magnetically breaking the surface region. On the other hand, the internal $T_{\phi z}$ stress will only transfer angular momentum from the surface to the disk midplane, and thus it won't lead to the overall disk accretion. The overall disk accretion is led by the $T_{R\phi}$ stress within the disk and the $T_{\phi z}$ stress at the disk atmosphere (e.g. the magnetocentrifugal wind). The detailed analysis on the surface accretion can be found in \cite{ZhuStone2018}. The accretion mechanisms are very similar. The only difference we notice by comparing Figure \ref{fig:onedrfig3rad500v2strongpaperb} in this work with Figure 7 in \cite{ZhuStone2018} is that $T_{\phi z}$ plays a more important role
in FU Ori disks which are thicker than disks in \cite{ZhuStone2018}. We have verified that the radiation viscosity is not important here. It is at least  5 orders of magnitude lower than the magnetic stress, which is
different from the sub-eddington accretion disks around supermassive black holes \citep{Jiang2019b}.

Although it is mainly the magnetic field that determines the accretion process, the radiation pressure in FU Ori plays a role in supporting the disk.
The lower panel of Figure \ref{fig:balance} shows the force balance with various terms in the vertical momentum equation (Equation \ref{eq:fullequation}).
In a steady state, the stress tensor divergence and the vertical gradient of the total pressure are balanced by the
vertical component of the gravitational force and the radiation pressure force. For a slowly moving fluid, the radiation pressure force is
$-{\bf S_r(P)}=\sigma_{t}{\bf F_{r,0}}/c$. Close to the disk midplane (the white region around z=0), it is mainly the  gradient of the gas pressure (the red curve) that balances the vertical forces (the black curves).
The magnetic pressure gradient (the blue curve) has the same strength as the radiation pressure (the black dotted curve, $\sim30\%$ of the gas pressure), and thus they balance each other. 
The stress tensor also contributes to compressing the disk. In the surface accreting region,  It is mainly the gradient of the magnetic pressure that balances
the gravity. Both the radiation pressure and the gradient of the gas pressure are negligible at the surface in comparison. This again suggests that the surface accretion occurs
in the magnetically dominated region. 

\subsection{Energy Budget}
Angular momentum transport and energy transport are the two most important aspects of accretion disks. 
In \cite{ZhuStone2018}, we have done analyses on the angular momentum budget of accretion disks threaded by net vertical magnetic fields. 
With the radiative transfer included in this work, we will do similar analyses for the disk's energy budget. 
The formulas are laid out in \S 2. Since the energy budget is related to the angular momentum budget, we will first repeat
the angular momentum analysis as we did in  \cite{ZhuStone2018}. 

The angular momentum budget is shown in the upper panel of Figure \ref{fig:onedconfig13full2sin43w0p98papervert}. 
Four different terms in the angular momentum equation (Equation \ref{eq:angsph}) are plotted. 
The $m_{r\phi}$ term is the radial gradient of the  $r$-$\phi$ stress (the first term on the right hand side of Equation \ref{eq:angsph}). 
After the integration over a volume in the disk, this term represents the transport due to the internal stress exerted at  the face that is perpendicular to the disk midplane,
 either from the turbulent stress or the stress due to the large scale organized magnetic fields. 
The $m_{\theta\phi}$ term is the $\theta$ gradient of the $\theta$-$\phi$ stress (the third term on the right hand side of Equation \ref{eq:angsph}). 
After the integration over a volume, it is the stress that is exerted at the disk surface. That is normally
due to the magnetocentrifugal disk wind. The other two terms
(the $\dot{m}_r$ term, which is the second term on the right hand side of  Equation \ref{eq:angsph}, and
the $\dot{m}_\theta$ term, which is the forth term on the right hand side of  Equation \ref{eq:angsph}) are the momentum transport due
to the radial and poloidal mass flux. In the thin disk theory, the poloidal mass flux term is small enough to be ignored so that the radial mass flux is balanced by
the $m_{r\phi}$ and $m_{\theta\phi}$  terms during the steady state.

In Figure \ref{fig:onedconfig13full2sin43w0p98papervert}, these terms are integrated
over $\theta$ from $\theta$=0.59 to 2.55 covering both the surface accreting region and the midplane region. Similar to the results in
\cite{ZhuStone2018}, the wind stress ($m_{\theta\phi}$) plays a less important role in accretion than the  $r$-$\phi$ stress. The $m_{\theta\phi}$ term
is $\sim$1/4 of the $m_{r\phi}$ term around $R\sim$1. Thus, only 20\% of accretion is due to the $\theta$-$\phi$ stress. On the other hand, this value is larger than 5\%
in the simulation of \cite{ZhuStone2018}. Considering that this disk is thicker than the disk in \cite{ZhuStone2018}, it implies that wind plays
a more important role for accretion in thicker disks. Nevertheless, most accretion is still due to the internal $r$-$\phi$ stress within the disk, as in \cite{ZhuStone2018}. 

On the other hand, the disk wind seems to play a much more important role in the energy transport. $Flux_r$, $Flux_\theta$, $E_{cool}$, and $E_{pot}$ in the lower panel of 
Figure \ref{fig:onedconfig13full2sin43w0p98papervert} are the
four terms on the right hand side of Equation \ref{eq:ensph}. The traditional thin disk theory (Equation \ref{eq:sphcool}) suggests that, far away from the inner boundary,
the energy transport in the radial direction (the first two terms on the right hand side of Equation \ref{eq:sphcool}) actually adds the disk energy by an amount that is equal to half the released gravitational energy. The energy gain/loss in the
poloidal direction is normally ignored. Thus, the total cooling rate is
1.5 times the released gravitational potential energy. However, our particular simulation suggests that energy transport in the radial direction (the red curve) is small compared with
the energy loss in the poloidal direction by the wind (the blue curve). The wind carries half of the gravitational potential energy (the green curve) 
so that only the rest half gravitational potential energy needs to be radiated away (the cyan curve).
Thus, the cooling rate is 
\begin{equation}
\langle Q_{cool}\rangle=\frac{\dot{\widetilde{M}}v_{K}^2}{2r^2}\,,
\end{equation}
which is roughly 1/3 of the value in the thin disk theory.
This cooling rate is plotted as the green dashed curve in the lower panel of Equation \ref{eq:ensph}, and it agrees with simulations  very well (even at the inner disk close to the inner boundary). 
Thus, the disk's effective temperature in the simulation can be approximated by 
\begin{equation}
\sigma T_{eff}^4=\frac{GM\dot{M}}{8\pi R^3}\,.\label{eq:teffsim}
\end{equation}

Based on our simulations,
such temperature estimate indeed agrees with the measured temperature at the $\tau_R\sim1$ surface.
The disk vertical structure at $R=0.1$ au is shown in Figure \ref{fig:Tstru}. At $\tau_R=1$ (the dotted line in the right panels), the value calculated using 
Equation \ref{eq:teffsim} (the blue curve in the temperature panel) agrees with the measured temperature very well. 

However, except for the similar $T_{eff}$, the temperature structure along $z$ in simulations is very different from the temperature structure based on the 
analytical theory. First, the radiation flux in the $\theta$ direction deviates significantly from the flux in the $z$ direction when $\tau_R\lesssim$1 (the bottom panels in Figure \ref{fig:Tstru}). 
This is because the radiation from the inner disk ($R<R_0$) is so strong that the flux measured in the optically thin region at $R_0$ consists of a significant contribution from the disk inside $R_0$. 
Thus, we use the measured flux at $\tau_R\sim$1 to represent the flux emitted by the local annulus at $R$. 
Second, the measured flux in either the $z$ direction or the $\theta$ direction rises much slower from the midplane to the $\tau_R=1$ surface than the models (red and blue solid curves) where the heating rate is proportional
to the disk local density (Equation \ref{eq:tauana}). The measured radiative flux only rises quickly beyond one disk scale height. This difference is due to: 
1) energy transport  by turbulence is as important as the radiative energy transport within the disk so that less temperature gradient is needed to radiate the thermal energy,
 as shown in the upper panel of Figure \ref{fig:balance};
2) both heating and accretion processes becomes more efficient at high above the disk midplane in our MHD simulations. 
Even with the similar emergent flux, the midplane temperature of the analytical $\alpha$ disk model is hotter than 
the measured midplane temperature in simulations by a factor of $\gtrsim$3. 
This result is consistent with previous local radiation MHD simulations \citep{Turner2004,Hirose2006,Jiang2014a}, suggesting that, towards the disk surface, MHD heating becomes more efficient compared with heating in viscous models. Third, the emergent flux at $\tau_R=1$ is significantly lower than
the flux (red curves) estimated based on the traditional accretion disk theory (Equation \ref{eq:Qcool}) using the measured disk accretion rate of $4\times10^{-4}\msunyr$. This is mostly
due to the energy lost in the poloidal direction as discussed in Figure \ref{fig:onedconfig13full2sin43w0p98papervert}. Equation \ref{eq:teffsim} which has accounted for the energy loss in the poloidal direction agrees with the measured $F_{z}$ at $\tau=1$ much better. We note that Equation \ref{eq:teffsim} only stands
at the inner disk. As shown in the temperature panel of Figure \ref{fig:onedradialfig2rad125}, the measured disk temperature is higher than the dotted line beyond $R\sim$0.2 au.
This is probably due to the fact that the outer disk is irradiated by the inner disk so that it gets heated up.

\subsection{Different Field Strengths and Geometries}
Since the disk temperature structure is self-consistently determined by the radiative transfer process in these simulations, 
the only major disk parameters that we can vary are the initial field geometry and strength. 
Thus, we carry out two additional simulations (V1e4 and T100) to explore how a weaker field or a toroidal field can affect the disk accretion.

The disk temperature, density, velocity, and magnetic field structures are shown in Figure \ref{fig:twodrhovelB600}. 
Although these two simulations have similar temperature structures, 
one major difference which is quite noticeable in the middle panels is that disk wind fails to be launched in the net toroidal field simulations. 
In T100, disk material high above the atmosphere falls to the disk (green curves) instead of leaving the disk. Furthermore, the surface accreting region in T100 is much thinner
if it exists at all. In the right panels, V1e4 shows an extended surface accreting region with high $B_{\phi}$ and $B_{r}$ values due to the surface accretion mechanism, while
T100 only shows a thin region at the disk surface with noticeable $B_{\phi}$ and very weak fields above that. There is no large-scale organized fields
in T100 either. The disk is dominated by turbulent fields in T100. 

This lack of surface accretion in net toroidal field simulations is also evident in Figure \ref{fig:onedradialfig2rad550compare} where the radial profiles  of various quantities are shown. 
In the $T_{total}$ and $\alpha$ panels, the two simulations have similar values at the disk midplane for both $T_{R\phi}$ and $\alpha$, 
while the vertically integrated $T_{R\phi}$ and $\alpha$ are significantly higher for V1e4.
This indicates that V1e4 has a higher stress level at the disk atmosphere than that in  T100. 
The magnetic field panel also shows that, while $B^2$   at the midplane is similar between two simulations, V1e4 has much stronger fields at the disk atmosphere. 
This leads to a higher accretion rate for V1e4 even though these two simulations have very similar turbulent levels at the disk midplane. 

The difference in disk wind is clearly shown in 
the vertical profiles of various quantities (Figure \ref{fig:onedrfig3rad600v2strongcomparepaperb}). 
At the wind region above $Z\sim R$, V1e4 has a much higher density than T100. The outflow nature of this region in V1e4 is clearly shown in the velocity panels, while this region in T100
is falling back to the disk. The magnetic fields and stresses are also very weak in the wind region of T100. Although there are some hints of surface accretion for T100 at $z/0.1$ au$\sim 1$ shown in the $v_R$ panel, the density there is more than 5 orders of magnitude lower than the disk midplane (the $\rho$ panel) so that the radial accretion of this surface is negligible in T100. 

Since net poloidal magnetic fields are essential for wind launching, it is important to understand how FU Ori's inner disk acquires such strong poloidal fields (tens to hundreds of Gauss). 
Current disk theory suggests that net poloidal magnetic fields can be either from the central star's magnetosphere, or inherited from the natal molecular cloud core. \cite{Konigl2011} have carried out 
MHD simulations to study how the kG magnetosphere of FU Ori's central star can interact with the fast accreting inner disk. They found that the magnetosphere truncation radius is pushed
close to the central star, but the wind that is launched at the truncation radius is still largely consistent with the observed outflow properties (e.g. mass loss rate and speed). On the other hand, the detailed modeling 
for wind lines  \citep{Calvet1993,Milliner2019} suggests that the wind is launched from a much larger scale (disk wind).
Thus, detailed synthetic observations for the simulations of  \cite{Konigl2011} are needed
to test if these simulations are consistent with the observed line profiles. For the second scenario, inheriting magnetic fields from molecular cloud cores has been studied extensively for disks controlled by both ideal MHD and non-ideal MHD processes \citep{RothsteinLovelace2008, GuiletOgilvie2012, GuiletOgilvie2013, Okuzumi2014, BaiStone2017} . Based on the simple field diffusion equation, the thin disk
can lose the magnetic fields outwards quickly \citep{Lubow1994}. But recent MHD simulations  by \cite{ZhuStone2018} found that the disk is quite thick for the perspective of the magnetic field structure, and the disk can actually transport field inwards slowly with time. Thus, FU Ori may gain poloidal magnetic fields from the outer disk during the low accretion state while material is piling up at the inner disk. When MRI is triggered
at the disk midplane \citep{Armitage2001,Zhu2009c}, such strong fields lead to strong accretion.

\begin{figure*}
\includegraphics[trim=0mm 50mm 0mm 1mm, clip, width=6.6in]{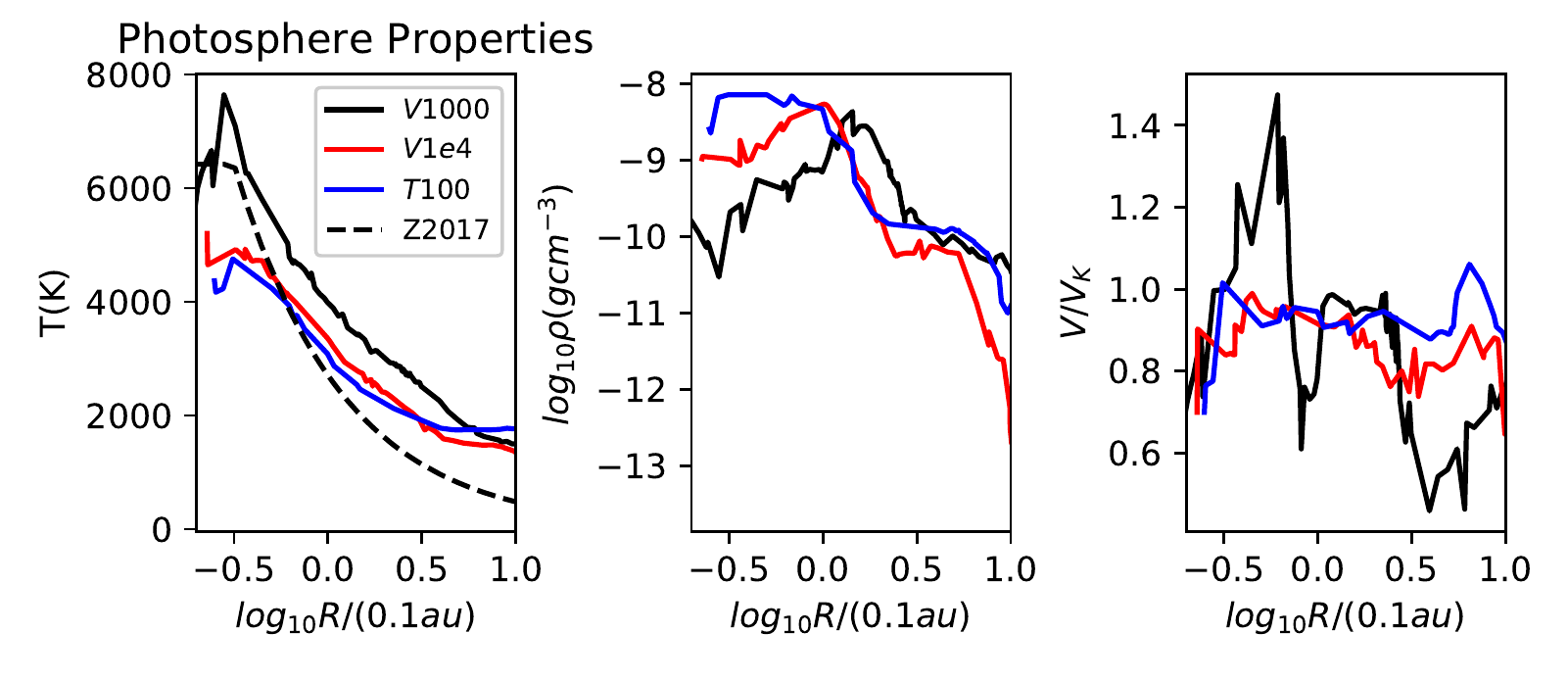}
\caption{ The temperature, density, and azimuthal velocity at the disk photosphere ($\tau_{R}$=1) along $R$ for three simulations. The dashed curve in the left panel is the effective temperature derived from FU Ori's SED modeling \citep{Zhu2007}. \label{fig:photosphere}}
\end{figure*}

\begin{figure*}
\includegraphics[trim=0mm 5.mm 0mm 1mm, clip, width=6.in]{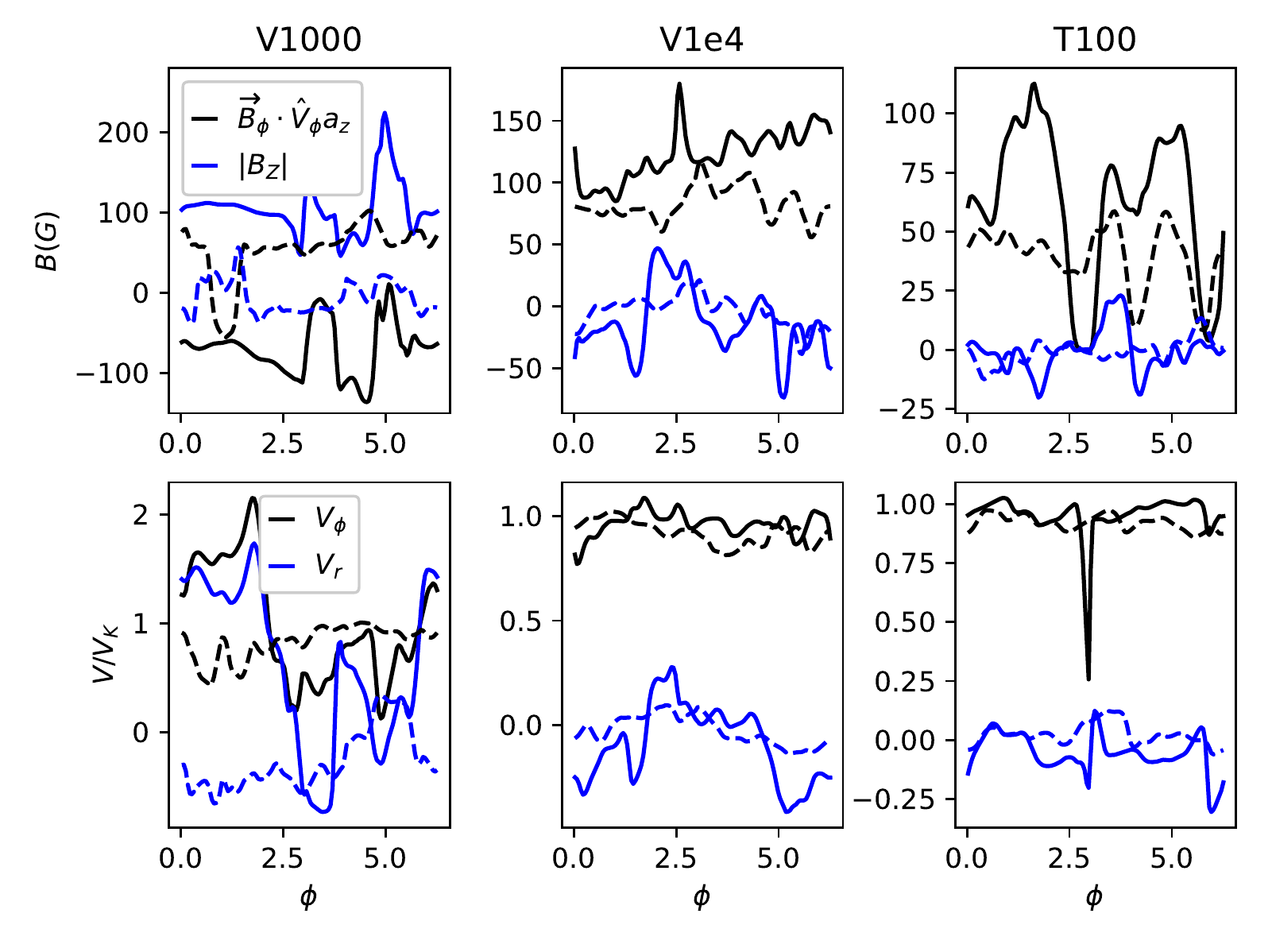}
\caption{Upper panels: the vertical (blue curves) and azimuthal (black curves) components of magnetic fields measured at the $\tau_R=1$ surface 
at $R$=0.05 au (solid curves) and 0.1 au (dashed curves) for three simulations (from left to right panels). 
$a_{z}$ equals 1 if the $B_{z}$ field at the $\tau_R=1$ surface is pointing in a direction that is leaving the disk midplane
and equals -1 if the $B_{z}$ field is pointing towards the midplane. $\hat{V}_{\phi}$ is the unit vector in the disk's rotational direction, 
and $\vec{B}_{\phi}$  is the projection of  the magnetic field vector to the disk's rotational direction. 
Lower panels: the radial (blue curves) and azimuthal (black curves) velocity at  the $\tau_R=1$ surface 
at $R$=0.05 au (solid curves) and 0.1 au (dashed curves) for three simulations.
\label{fig:bfieldtau1paper500}}
\end{figure*}

\begin{figure}
\includegraphics[trim=110mm 125mm 120mm 1mm, clip, width=2.7in]{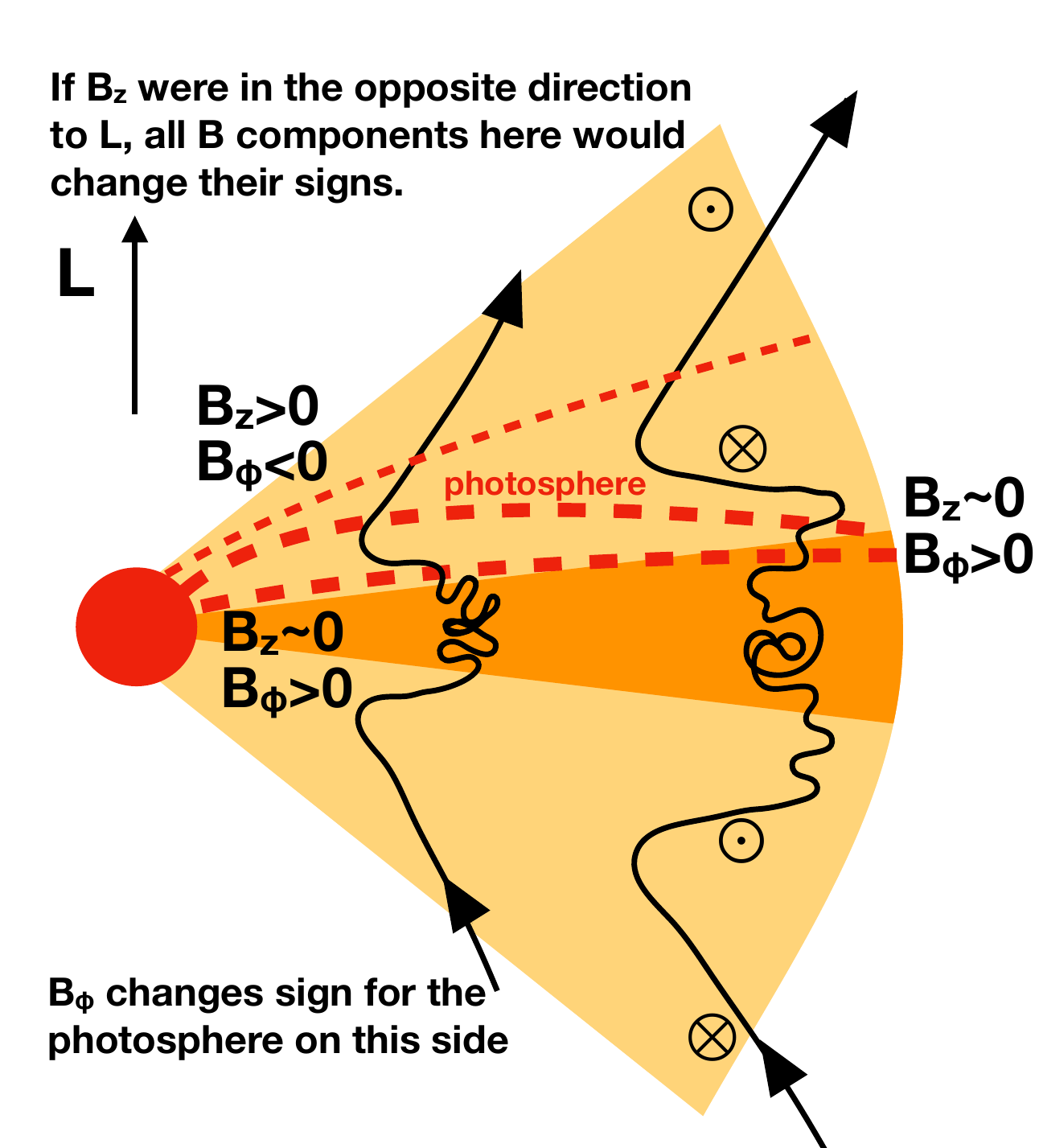}
\caption{The schematic plot showing $B_z$ and $B_{\phi}$ measured at different radii under different scenarios (the thin red curve: the photosphere is always at the wind region;
the middle thick red curve: the photosphere of the inner disk is at the wind region while the photosphere at the outer disk is within the disk; the lower thick red curve: the photosphere is always
within the disk region) .  
The signs of $B$ follow the right hand rule with respect to the angular momentum vector of the disk.
\label{fig:scheme}}
\end{figure}

\begin{figure*}
\includegraphics[trim=0mm 5.mm 0mm 1mm, clip, width=6.6in]{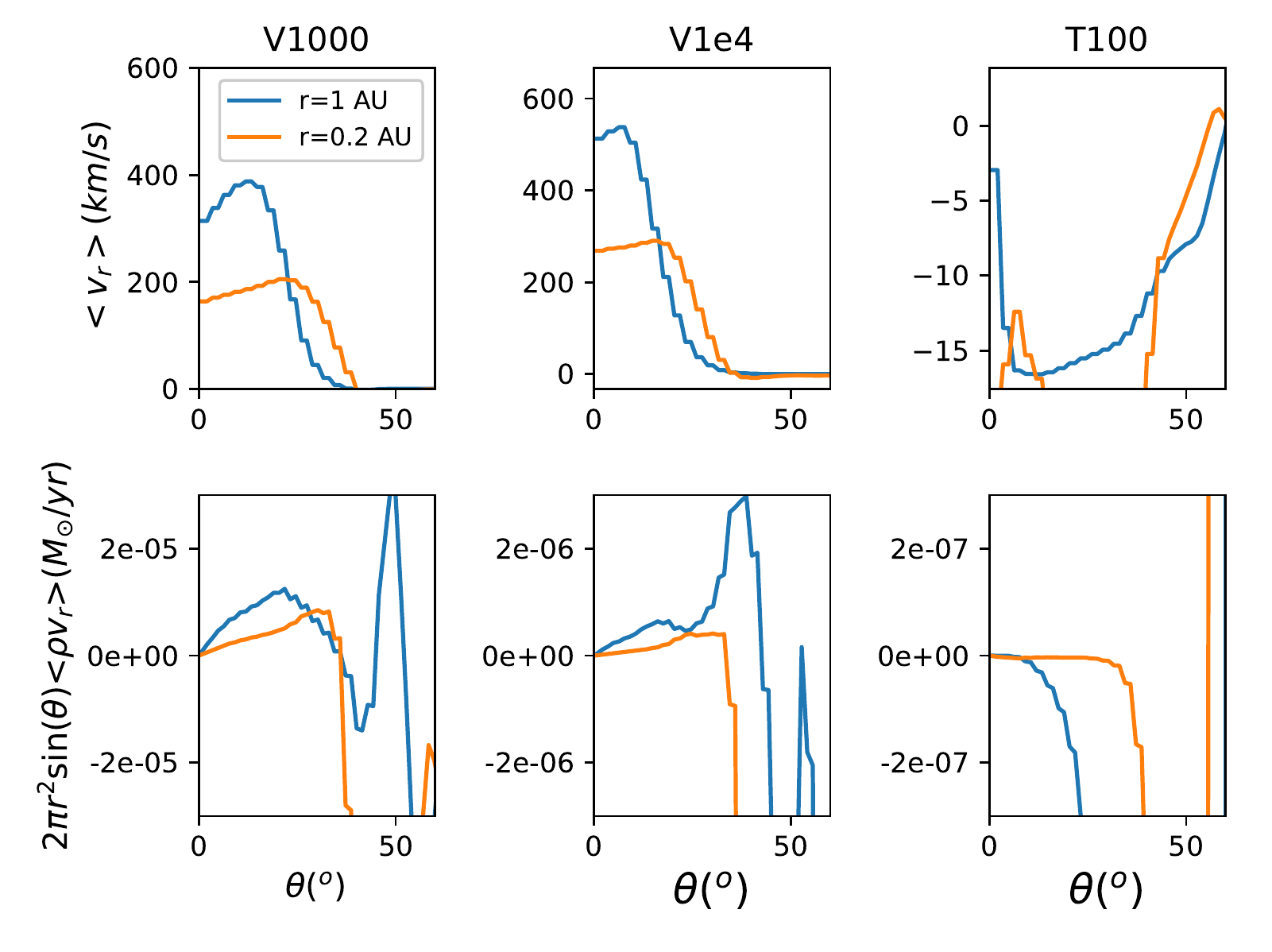}
\caption{The radial velocity (upper panels) and mass loss rate (lower panels) at 0.2 au and 1 au along the $\theta$ direction in our three simulations (from left to right). 
The quantities have been averaged over both time (the last 2$T_{0}$ of each simulation) and azimuthal direction. \label{fig:outerboundarytorfield}}
\end{figure*}

\section{Discussion}
After studying the disk's physical structure, we will compare the simulations with existing observations regarding the disk temperature, magnetic fields and disk wind.

\subsection{Photosphere Properties}
 Previous FU Ori SED modeling from \cite{Zhu2007} suggests that the disk's effective temperature follows the standard viscous disk model and the disk's maximum effective temperature is $\sim$6420 K.  This temperature profile is plotted against the photosphere temperature  (at $\tau_R$=1) in our simulations, shown in Figure \ref{fig:photosphere}. Our fiducial model (V1000) has a similar
maximum disk temperature as the observations, although its accretion rate ($\sim5\times10^{-4}\msunyr$) is twice the accretion rate used in \cite{Zhu2007} (2.4$\times10^{-4}\msunyr$).  Considering that most disk luminosity comes from the hottest region, our fiducial model has a similar luminosity as the observation. All our simulations
have flatter profiles compared with observations, which is due to the irradiation from the inner disk to the outer disk as discussed above. Thus, our simulations may need to be combined with a slightly different
extinction curve from \cite{Zhu2007} to explain all the observations at different wavelengths. The photospheres in our simulated disks have densities of 10$^{-10}$-10$^{-9}$ g cm$^{-3}$, and almost rotate at the local Keplerian speed.  

\subsection{Comparison with Magnetic Field Zeeman Observations}
\cite{Donati2005} use a high resolution spectropolarimeter to measure circularly polarized light (Stokes $V$) from thousands of spectral lines for FU Ori. 
The circular polarized light is produced by Zeeman splitting which depends on both the field geometry and strength. 
The measured polarization signal corresponds to the line-of-sight magnetic field of $\sim$ 32 G. Together with some additional constraints on the disk parameters (e.g. 60$^o$ inclination) and
theoretical disk wind models \citep{Ferreira1997}, 
the detailed decomposition of the Stokes $V$ into antisymmetric and
symmetric components has put a much more stringent constraint on the magnetic fields of FU Ori. To summarize the findings: 1)  comparing the polarized light with the unpolarized light reveals
that strong magnetic fields occupy $\sim20\%$ of the disk surface, and the magnetic plasma rotates $\sim$2-3 times slower than the local Keplerian velocity; 2) the vertical
component of the magnetic fields (leaving the disk surface) is pointing towards us with a strength of $\sim$1 kG at 0.05 au; 3) the toroidal fields in the disk point to a direction which is opposite to
the disk's orbital rotation with a strength of $\sim$ 500 G at 0.05 au. 

Although these measurements are consistent with previous resistive MHD simulations \citep{Ferreira1997} where the MRI turbulence is simplified by the resistivity parameters,
we can now compare these observations directly with our first-principle radiation MHD simulations. We thus measure the magnetic field direction and strength at the $\tau_R=1$ surface in our simulations. 
The magnetic fields at $R$=0.05 au and 0.1 au are shown in Figure \ref{fig:bfieldtau1paper500}. Please note the direction of the magnetic field in this figure. $a_{z}$ is a parameter that equals 1 if
$B_{z}$ at the $\tau_R=1$ surface is pointing in a direction that is leaving the disk midplane and it is -1 if $B_{z}$ is pointing towards the disk midplane.  $\hat{V}_{\phi}$ is the unit vector in the disk's rotational direction. The reason that we express $B_{\phi}$ in this 
$\vec{B}_{\phi}\cdot\hat{V}_{\phi}a_z$ form is due to the facts that we can view the disk from either the top or bottom side of the disk in Figure \ref{fig:twodrhoBstrong} and the disk's $B_{z}$
can also be either aligned or  anti-aligned with the angular momentum vector of the disk's rotation. Let's take the V1000 case as an example. As shown in the upper left panel of 
Figure \ref{fig:bfieldtau1paper500},  $\vec{B}_{\phi}\cdot\hat{V}_{\phi}a_z$ (the solid black curve) is negative. 
If we observe the disk downwards from the upper side of the disk in Figure \ref{fig:twodrhoBstrong}, $B_{z}$ is pointing
to us so that $a_{z}=1$. In this case $\vec{B}_{\phi}\cdot\hat{V}_{\phi}$ is negative implying that $B_{\phi}$ is in the opposite direction from the disk rotation. This can be seen in Figure \ref{fig:twodrhoBstrong} where $B_{\phi}$ has negative values in the wind region. If we view the disk from the bottom and $\vec{B}_{z}$ is pointing towards the disk midplane, $a_{z}=-1$ so that
$B_{\phi}$ at the $\tau_R=1$ surface on this side of the disk is in the same direction as the disk rotation (as shown with the positive $B_{\phi}$ values at the bottom side of the wind region 
in Figure \ref{fig:twodrhoBstrong}). On the other hand, since we don't know if the rotational axis of the disk is aligned or anti-aligned with the magnetic fields (e.g. both Sun and Earth have magnetic reversals),
we can reverse the field direction in simulations  and the disk velocity structure will be unchanged.  In that case, if we look at the disk downwards from the upper side of Figure \ref{fig:twodrhoBstrong}, $a_{z}=-1$ and $B_{\phi}$ at the wind
region will be positive (in the same direction as the disk rotation) so that $\vec{B}_{\phi}\cdot\hat{V}_{\phi}a_{z}$ is still negative.

Our fiducial case (V1000) roughly reproduces the velocity and field geometries inferred from \cite{Donati2005}. At R=0.05 au, the $\tau_R=1$ surface is at $z\sim R$ which is the top of the surface accreting region or the bottom of the wind region  (Figure \ref{fig:twodrhoBstrong}). At $z\sim R$, the disk rotates with $\sim$60\% of the midplane Keplerian velocity (the lower left panel of Figure \ref{fig:bfieldtau1paper500}), while the disk becomes Keplerian slightly deeper in the disk (the $V_{\phi}$ panel in Figure \ref{fig:onedrfig3rad500v2strongpaper}). Considering that the photospheres in other two cases are slightly deeper and they are Keplerian rotating, this $\sim$60\% of Keplerian rotation speed sensitively depends on the photosphere position and can be quite uncertain. 
At the $\tau_R=1$ surface of R=0.05 au, the field strength is quite strong with $B_{z}\sim 150$ G. If $B_{z}$ is pointing to us, $B_\phi$ will be in a direction that is opposite to the disk rotation, which is consistent with observations. $B_{\phi}$ is half of $B_{z}$, which is also consistent with observations.  At deeper regions in the disk, both $B_{\phi}$
and $B_{z}$ decreases significantly. In the surface accreting region and down towards the disk midplane, $B_{\phi}$ changes from negative to zero and to positive. Thus, the 20\% covering factor from observations could be that 20\% light comes from
the strong $B$ and sub-Keplerian region, while the rest 80\% comes from the deeper Keplerian and weaker $B$ region. The only difference between our simulations and the observations is that
the field strength measured in simulations is weaker than the observed inferred kG strength by a factor of $\sim$5.  On the other hand, we note that the 
first-order moment of the observed Zeeman signature is only $\sim$ 32 G. The kG strength is inferred from matching models considering the $60^o$ 
inclination and the assumed filed geometry and filling factor.
As will be shown in Section 5.3, the assumed inclination is too high compared with recent ALMA observations. Overall, the relatively good agreement regarding the field and 
velocity structure is very encouraging. 

Our model also predicts that
new observations by SpIROU at near-IR may reveal a different field structure than earlier results using optical lines from \cite{Donati2005} since near-IR lines come from further out in the disk (e.g. 0.1 au). 
The simulation indicates that the $\tau_R=1$ surface has very different field geometries and strengths at $R=0.1$ au (the dashed curves in Figure \ref{fig:bfieldtau1paper500}) compared with those at $R=0.05$ au. From Figure \ref{fig:twodrhoBstrong}, we can see that, further away from
the central star, the $\tau_R=1$ surface is closer to the disk midplane due to the lower disk surface density there. The upper left panel in Figure \ref{fig:onedrfig3rad500v2strongpaper} shows that
both $B_{z}$ and $B_{\phi}$ at the $\tau_R=1$ surface change their signs moving from 0.05 au to 0.1 au and the field strength gets a lot weaker. Furthermore, unlike at 0.05 au, $B_{\phi}$ is stronger than $B_{z}$ at the photosphere of 0.1 au since the photosphere is at the bottom of the surface accreting region and closer to the disk midplane. 

The surface accreting regions in our other two simulations, V1e4 and T100, have much lower density so that the $\tau_R=1$ surface is close to the disk 
midplane even at $R=0.05$ au (Figure \ref{fig:twodrhovelB600}).  Thus, $B_{\phi}$ is always stronger than $B_{z}$ at the photosphere 
as shown in the right two panels of Figure \ref{fig:bfieldtau1paper500}.
If $B_{z}$ is pointing towards us, $B_{\phi}$ will be in the same direction as the disk rotation in these cases. 

Various possible scenarios for $B_{z}$ and $B_{\phi}$ measurements are summarized in Figure \ref{fig:scheme}. Under the surface accretion picture, $B_{z}$ becomes quite strong at
the upper surface/the base of the wind region at $R\sim z$, and $B_{\phi}$ changes sign there. Thus, if the disk has a very high density so that the photosphere is only in the wind region or  at the wind-base region (the thin dashed curve is the photosphere under this scenario), we
are expecting to measure strong $B_{z}$ and $B_{\phi}$ at all disk radii. On the other hand, the disk normally has a lower density at the outer cooler region and the opacity there is lower, it is more likely that the photosphere changes from the wind-base region at the inner disk to the lower surface/disk region at the outer disk (e.g. V1000 case). In this case,  the $B_{z}$ at the photosphere decreases 
dramatically at the outer disk and $B_{\phi}$ changes sign
from the inner photosphere to the outer disk photosphere, indicating observations at different wavelengths may reveal different field and velocity geometries. 
For the third scenario that the photosphere is always closer to the disk (e.g. V1e4 and T100 cases), $B_{z}$
will be significantly smaller than $B_{\phi}$ at all radii and observations at different wavelengths may reveal similar field and velocity geometries. 
We note that the signs of various $B$ components can change depending on our viewing angle and the orientation between the fields and the rotational axis (as described in Figure \ref{fig:scheme}).

 We want to caution that we use the $\tau_R$=1 surface to represent both the photosphere and where the magnetic fields are measured. In reality, the magnetic fields are measured by \cite{Donati2005} using
a subset of G0  line list. These lines are likely to trace disk region that is above the photosphere. Detailed radiative transfer modeling with lines is needed to compare our simulations with observations.

\subsection{Comparison with Disk Wind Observations}
FU Ori shows evidence of strong winds in P Cygni profiles, especially in the Na I resonance lines \citep{Bastian1985,Croswell1987}. 
The blue-shifted line absorption implies a disk outflow with a typical velocity of 100-300 km/s and a mass loss rate of $\sim$10$^{-5}\msunyr$ \citep{Calvet1993}.
Recent work by \cite{Milliner2019} suggests that the wind may be turbulent. 

We have plotted the gas radial velocity and mass loss rate at different poloidal directions in Figure \ref{fig:outerboundarytorfield}.
As long as the disk is threaded by net vertical fields, the magnetic fields accelerate the gas flow along the radial direction, reaching $\sim$400 km/s
terminal velocity. The integrated outflow rate at a distance $r$ from the central star is
\begin{eqnarray}
\dot{M}_{wind}(r)&=&\int_{0}^{2\pi}d\phi\int_{0}^{\pi}d\theta r^2\sin(\theta)\rho v_{r}\\
&=&\int_{0}^{\pi}2\pi r^2\sin(\theta)\langle\rho v_{r}\rangle d\theta \,,
\end{eqnarray} 
where $\langle\rangle$ means that the quantities have been averaged over the azimuthal direction. 
The lower left panel of Figure  \ref{fig:outerboundarytorfield} shows that $2\pi r^2\sin(\theta)\langle\rho v_{r}\rangle$ is around $10^{-5}\msunyr$.
Thus, the integrated wind loss rate  from the pole to 30$^o$ (0.52 in Radian) away from the pole is
$\sim10^{-5}\msunyr*0.52*2\sim10^{-5}\msunyr$ where 2 comes from both sides of the disk. Thus, our  fiducial simulation 
can reproduce both the observed outflow velocity and outflow rate.

If the disk is threaded by net toroidal fields, wind can not be launched, as shown in the right panel of Figure \ref{fig:outerboundarytorfield}.
Thus, the existence of disk wind in FU Ori implies that the disk is threaded by net vertical magnetic fields. 

\subsection{New FU Ori Parameters}

\begin{figure*}
\includegraphics[width=6.6in]{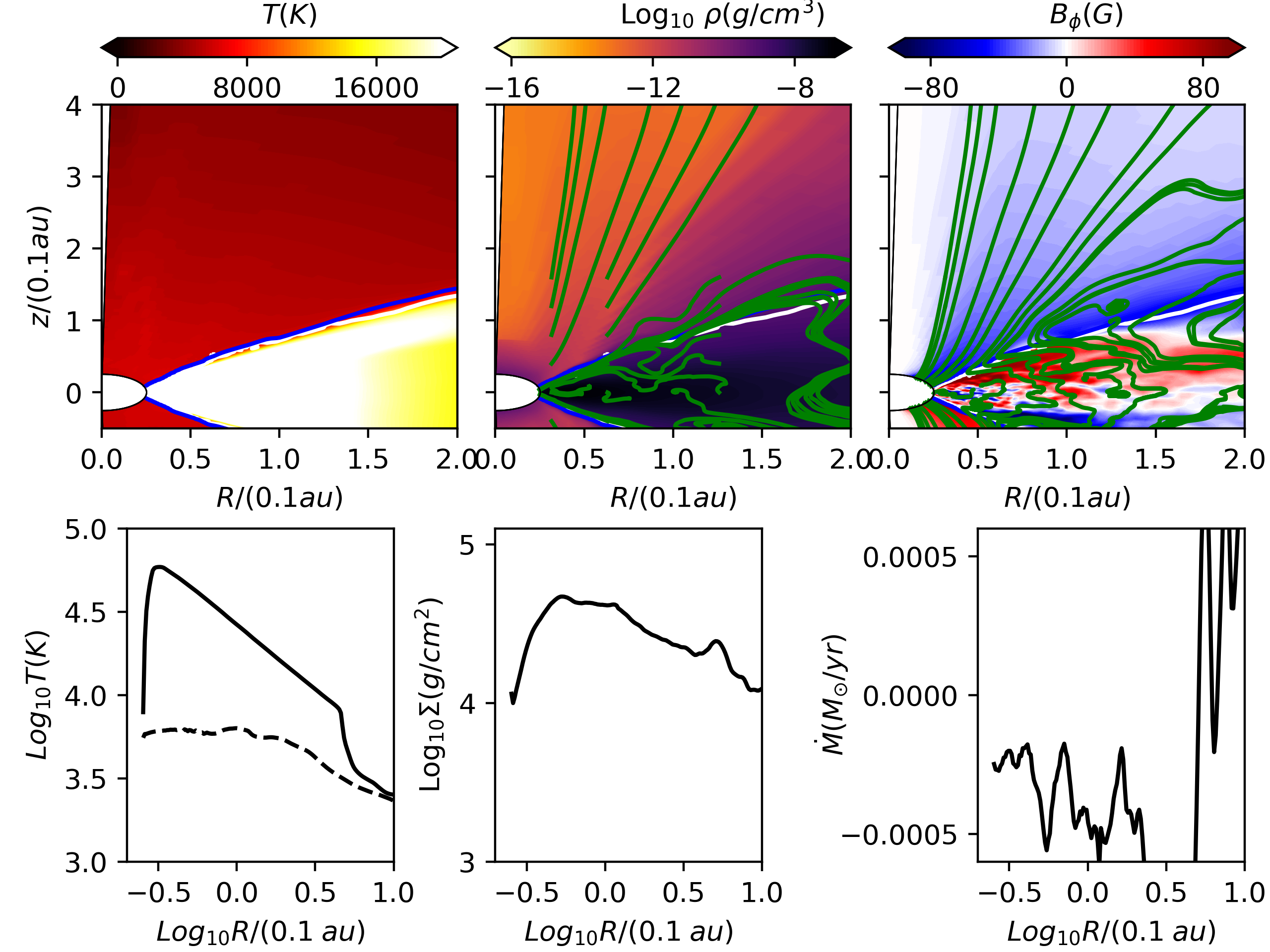}
\caption{Similar to Figure \ref{fig:twodrhoBstrong} and Figure \ref{fig:onedradialfig2rad125} but for the new FU Ori parameters at $t=31.5 T_{0}$. 
In the lower left panel, the dashed curve is the temperature at $\theta=0.9$ where the photosphere is. \label{fig:twodrhovelBM0p6}}
\end{figure*}

While we are preparing this manuscript, the distance to FU Ori is more precisely constrained by Gaia. The new distance is 416$\pm$9 pc  \citep{Gaia2018} instead of 500 pc assumed in \cite{Zhu2007}.
The disk inclination is also better constrained to be 35$^o$ by ALMA \citep{Perez2019} instead of 55$^o$ assumed in \cite{Zhu2007}. With these updated parameters, \cite{Perez2019} derive that
the central star mass is updated to be 0.6 $M_{\odot}$ instead of 0.3 $M_{\odot}$, and the disk accretion rate is 3.8$\times 10^{-5}\msunyr$ instead of 2.4$\times 10^{-4}\msunyr$. The disk accretion rate
now is only 1/6 of the earlier estimate due to the fact that both the closer distance and more face-on configuration reduce the disk accretion rate estimate. In the Appendix and Figure \ref{fig:newfuorised}, we have shown the SED fitting using the new parameters. 

To be consistent with these new parameters, we have carried out a simulation which is similar to the V1e4 case but with $M_{*}=0.6 M_{\odot}$. The results are shown in Figure \ref{fig:twodrhovelBM0p6}.
The overall ``surface accretion'' picture still stands. But due to the short duration of this simulation (only to 31.5 $T_0$), the field structure at the surface accreting region is not fully established. 
The high disk accretion rate and the high central star mass release a significantly amount of gravitational energy so that the disk is significantly hotter than the V1e4 case with $M_{*}=0.3 M_{\odot}$.
The real FU Ori system may have weaker net vertical fields or a lower surface density than those we assumed in Figure \ref{fig:twodrhovelBM0p6}.

\section{Conclusions}
We have carried out three-dimensional global ideal MHD simulations to study the inner outbursting disk of FU Ori.
Since the accretion disk outshines the central star, the radiation field of the disk plays an important role in the disk accretion dynamics. 
The radiative transfer is also crucial for connecting with observations. Thus, we self-consistently solve the radiative
transfer equations along with the fluid MHD equations. We have carried out simulations where 
the disk is threaded by either net vertical or net toroidal magnetic fields.

We find that, when the disk is threaded by net vertical fields,
most accretion occurs in the magnetically dominated atmosphere at z$\sim$R, very similar to the ``surface accretion'' mechanism
in previous simulations with the simple locally isothermal equation of state. This implies that the ``surface accretion'' is
a general feature of accretion disks threaded by net vertical fields. The disk midplane shows spiral arms while the disk surface has filamentary structures. 
With radiative transfer included, we can study the accretion disk's temperature structure.
The radiation pressure is $\sim30\%$ of the gas pressure at the inner disk (e.g. 0.1 au).
The disk midplane has a sharp temperature transition at $\sim$0.15 au separating the inner and outer disks which are at the higher and lower branches of the equilibrium ``S'' curve. 
But the accretion and stress profiles are smooth despite the jump of disk temperature. This implies that the global accretion structure is mainly
controlled by the global geometry of magnetic fields and is insensitive to the disk local temperature.  

Compared with the simulations for thinner disks in \cite{ZhuStone2018}, the simulations here have stronger disk wind. 
20\% of disk accretion is due to the wind $\theta-\phi$ stress, which is higher than 5\% in \cite{ZhuStone2018}.
The wind mass loss rate from the disk surface spanning one order of magnitude in radii is 1-10\% of the disk accretion rate, which is also higher
than 0.4\% in \cite{ZhuStone2018}. Thus, the disk wind seems to be stronger in thicker disks. 
The mass loss rate of $\sim$10$^{-5}\msunyr$ in our FU Ori simulations is consistent with observations.
The wind's terminal speed is $\sim$300-500 km/s.
This speed is also consistent with the observed wind speed and is several times the Keplerian speed at the launching point ($V_{K}$ at the inner boundary is 100 km/s). 
On the other hand, no disk wind is launched when the disk is threaded by net toroidal fields, implying that net vertical fields are crucial for launching the disk wind.
The net toroidal field simulation also shows weaker accretion and smaller vertically integrated stresses due to the lack of the surface accretion at the disk surface.

The moderate disk wind also carries half of the accretion gravitational potential energy so that only the rest half of gravitational potential energy needs to be radiated away.
The emergent flux is only $\sim$1/3 of the traditional value with the same disk accretion rate (comparing Equation \ref{eq:teffsim} with Equation \ref{eq:Qcoolteff}).  
Thus, the disk photosphere temperature is lower than that predicted by the thin $\alpha$-disk model having the same accretion rate.
Thus, using the observed flux, the previously inferred disk
accretion rate may be lower than the real disk accretion rate by a factor of  $\sim$2-3.   
The disk midplane is also much cooler than that predicted by viscous models due to  the energy transport by turbulence at the midplane and the efficient heating at the disk surface.
With the surface accretion, the disk is heated up at the surface and the energy there can be more easily radiated away. 

We have compared the magnetic fields at the photosphere in our simulations with Zeeman observations from \cite{Donati2005}.
The disk's $\tau_R=1$ photosphere can be either 
in the wind launching region or the accreting surface region, depending on the accretion rates and the disk radii. Magnetic fields
have drastically different directions and magnitudes between these two regions. 
It is very encouraging that the photosphere in our fiducial model, which is at the base of the wind launching region, agrees with previous Zeeman observations regarding both
the magnetic field direction and magnitude.
On the other hand, we suggest that the magnetic fields probed by future Zeeman observations at different wavelengths (e.g. near-IR) or for different systems (e.g. with lower accretion rates)
can be quite different from the existing measurements in \cite{Donati2005} since the photosphere can be deep into the surface accreting region.

Overall, we find excellent agreements between the first-principle MHD simulations having net vertical fields and existing observations regarding both the wind and magnetic field properties. This strongly supports
that accretion disks in FU Orionis systems are threaded by net vertical magnetic fields and MHD processes are important for the accretion process.
More comparisons between simulations and future observations will allow us to probe the 3-D structures of magnetic fields and gas flow in accretion systems.

\section*{Acknowledgments}
All  simulations are carried out using computer supported by the 
Texas Advanced Computing Center (TACC) 
at The University of Texas at Austin through XSEDE grant TG-
AST130002 and from the NASA High-End Computing (HEC) Program through the NASA Advanced Supercomputing (NAS) Division at Ames Research Center. 
Z. Z. acknowledges support from the National Science Foundation under CAREER Grant Number AST-1753168. 
The Center for Computational Astrophysics at the Flatiron Institute is
supported by the Simons Foundation.

\bibliographystyle{mnras}
\input{msMNRAS.bbl}

\appendix
\section{SED fitting for FU Ori}
\begin{figure}
\includegraphics[width=3.3in]{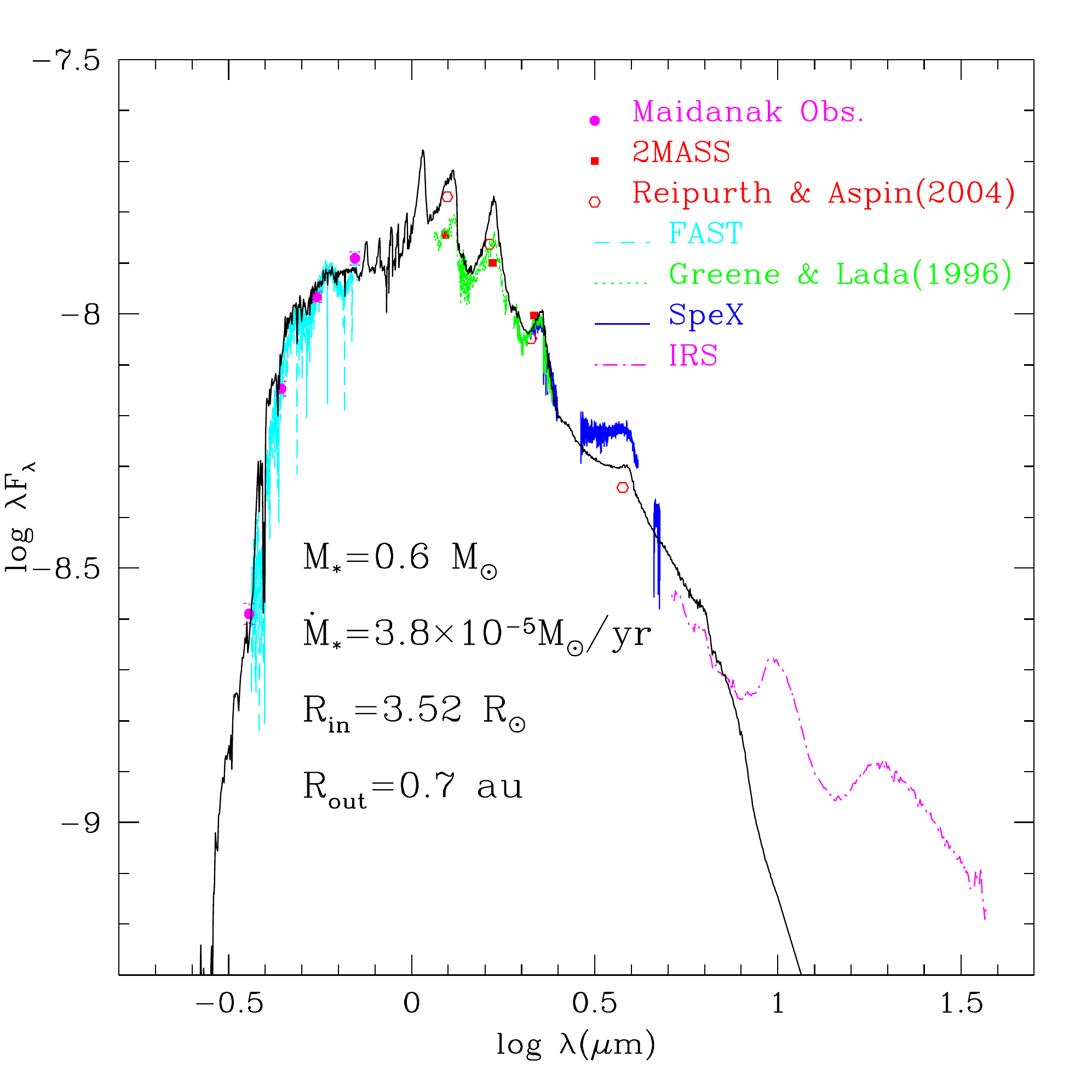}
\caption{With the new FU Ori distance from Gaia and disk inclination from ALMA, FU Ori's disk parameters have changed moderately \citep{Perez2019}. This shows the  new SED fit using the updated FU Ori parameters \citep{Perez2019}. \label{fig:newfuorised}}
\end{figure}

With the updated FU Ori inclination, \cite{Perez2019} use the disk atmospheric radiative transfer model \citep{Zhu2007} to update FU Ori's parameters.
The best fit SED is shown in Figure \ref{fig:newfuorised}.

\bsp
\label{lastpage}
\end{document}